\documentclass[preprint,12pt]{elsarticle}

\usepackage{amsmath}
\usepackage{amsfonts}
\usepackage{amssymb}
\usepackage{graphicx,psfrag,epsf}
\usepackage{enumerate}
\usepackage{natbib}
\usepackage{url} 
\usepackage{multirow}
\usepackage{color}
\usepackage{booktabs}
\usepackage{float}
\usepackage{caption}
\usepackage{subcaption}

\journal{Computational Statistics \& Data Analysis}

\begin{document}

\begin{frontmatter}




\title{FFT-Based Fast Bandwidth Selector for Multivariate Kernel Density Estimation \tnoteref{t1}}
\tnotetext[t1]{The up-to-date \textsf{R} source codes are included as a supplemental material. One can replicate all the figures and all the data shown in the tables. }


\author[ag]{Artur Gramacki\corref{cor1}}
\ead{a.gramacki@issi.uz.zgora.pl}

\author[jg]{Jaros\l{}aw Gramacki}
\ead{j.gramacki@ck.uz.zgora.pl}

\cortext[cor1]{Corresponding author} 

\address[ag]{
    Institute of Control and Computation Engineering,
    University of Zielona G\'o{}ra,
    ul. Licealna 9, Zielona G\'o{}ra 65-417, Poland
}

\address[jg]{
    Computer Center,
    University of Zielona G\'o{}ra,
    ul. Licealna 9, Zielona G\'o{}ra 65-417, Poland
}

\begin{abstract}
The performance of multivariate kernel density estimation (KDE) depends strongly on the choice of bandwidth matrix. The high computational cost required for its estimation provides a big motivation to develop fast and accurate methods. One of such methods is based on the Fast Fourier Transform. However, the currently available implementation works very well only for the univariate KDE and it's multivariate extension suffers from a very serious limitation as it can accurately operate only with diagonal bandwidth matrices. A more general solution is presented where the above mentioned limitation is relaxed. Moreover, the presented solution can by easily adopted also for the task of efficient computation of integrated density derivative functionals involving an arbitrary derivative order. Consequently, bandwidth selection for kernel density derivative estimation is also supported. The practical usability of the new solution is demonstrated by comprehensive numerical simulations.

\end{abstract} 


\end{frontmatter}

\section{Introduction}\label{sec-intro}
Kernel density estimation (KDE) is a very important statistical technique with many practical applications. It has been applied successfully to both univariate and multivariate problems. There exists extensive literature on this issue, including several classical monographs, see \cite{Silverman:1986}, \cite{Scott:1992} and \cite{Wand-1995}.

A general form of the $d$-dimentional multivariate kernel density estimator is
\begin{align}\label{eq-kde}
\hat{f}(\boldsymbol{x},\boldsymbol{H}) 
&=
\frac{1}{n} 
\sum_{i=1}^n 
K_{\boldsymbol{H}} \left( \boldsymbol{x}-\boldsymbol{X}_i \right),  
\end{align}	
where
\begin{align}\label{eq-Hu}
K_{\boldsymbol{H}}(u) 
&= 
|\boldsymbol{H}|^{-1/2} K \left(\boldsymbol{H}^{-1/2}u\right),
\end{align}
and
$\boldsymbol{H}$ is the $d \times d$ \emph{bandwidth} or \emph{smoothing} matrix, $d$ is the problem dimensionality,	$\boldsymbol{x}=(x_1, x_2, \ldots, x_d)^T$, and	
$\boldsymbol{X}_i=(X_{i1}, X_{i2}, \ldots, X_{id})^T$, $i=1,2,\ldots,n$ is a sequence of independent identically distributed (iid) $d$-variate random variables drawn from a (usually unknown) density function $f$. 
Here $K$ and $K_{\boldsymbol{H}}$ are the unscaled and scaled kernels, respectively. In most cases the kernel has the form of a standard multivariate normal density. 

The univariate kernel density estimator for a random sample $X_1, X_2, \ldots X_n$ drawn from a common and usually unknown density function $f$ is given by
\begin{align}\label{eq-kde-1D}
\hat{f}(x, h) 
=
\frac{1}{n} 
\sum_{i=1}^n 
K_h \left( x - X_i \right),  
\end{align}
where
\begin{align}\label{eq-hu}
K_h(u) 
= 
h^{-1} K \left(h^{-1}u\right),
\end{align}
and $h$ is the bandwidth which is a positive integer. The scaled ($K_h$) and unscaled ($K$) kernels are related in Eq.~(\ref{eq-hu}). Note that the notation used in Eq.~(\ref{eq-kde}) is not a direct extension of the univariate notation in Eq.~(\ref{eq-kde-1D}), since in the one-dimensional case the bandwidth is $\boldsymbol{H}=h^2$, so we are dealing with `squared bandwidths' here.

It seems that both uni- and multivariate KDE techniques have reached maturity and recent developments in this field are primarily focused on computational problems. There are two main computational problems related to KDE: (a) the fast evaluation of the kernel density estimates~$\hat{f}$, and (b) the fast estimation (under certain criteria) of the optimal bandwidth matrix $\boldsymbol{H}$ (or scalar $h$ in the univariate case). As for the first problem, a number of methods have been proposed, see for example \cite{Raykar:2010} for a comprehensive review. As for the second problem, relatively less attention has been paid in the literature. An attempt of using the Message Passing Interface (MPI) was presented in \cite{Lukasik:2007}. In \cite{Raykar:2006} the authors give an $\epsilon$-$exact$ approximation algorithm, where the constant $\epsilon$ controls the  desired arbitrary accuracy. Other techniques, like for example usage of Graphics Processing Units (GPUs), have also been used \citep{Andrzejewski:2013}.
In this paper we are concerned with fast estimation of the optimal bandwidth and are interested in the multivariate case only. However, our results can be easily adapted also to the univariate case.

It is obvious from Eq.~(\ref{eq-kde}) that the naive direct evaluation of the KDE at $m$ evaluation points for $n$ data points requires $O(mn)$ kernel evaluations. Evaluation points can be of course the same as data points and then the computational complexity is $O(n^2)$ making it very expensive, especially for large datasets and higher dimensions. 

As for finding of the optimal bandwidth, the computational problems are even more evident. Typically, to complete all the required computations for this task a sort of numerical optimization is needed. Usually, the computational complexity of evaluating typical objective function is $O(n^2)$. During the optimization process the objective function must be evaluated many times (often more that a hundred or so), making the problem of finding the optimal bandwidth very expensive, even for moderate data dimensionalities and sizes. 

In this paper we are concerned with an FFT-based method that was originally described in \cite{Wand-1994a}. In \cite[appendix D]{Wand-1995} an interesting illustrative toy example has been presented. From now on this method will be called \emph{Wand's algorithm}. It can be used for the KDE evaluation as well as for bandwidth selection problem and it works very well for the univariate case given in Eq.~(\ref{eq-kde-1D}). Unfortunately, its multivariate extension does not support \emph{unconstrained} bandwidth matrices (that is, if $\boldsymbol{H} \in  \mathcal{F}$, where $\mathcal{F}$ is the set of all symmetric, positive definite $d \times d$ matrices). The method supports only more restricted \emph{constrained} bandwidth matrices (that is, if $\boldsymbol{H} \in  \mathcal{D}$,  where  $\mathcal{D}$ is the set of all positive definite diagonal matrices of the form $\boldsymbol{H}=\mathnormal{diag}(h_1^2,\ldots,h_d^2)$). This limitation was successfully overcome by the authors and the main results are presented in \cite{Gramacki:2016}. 
In this paper we extend those results to the problem of fast (FFT-based) estimation of unconstrained bandwidth matrices. 

To the best of our knowledge, our paper is the first where this problem is presented and successfully solved using an FFT-based approach. In this work we use excellent results presented in \cite{Chacon:2015}, clearly the ones which significantly simplifies computations of integrated density derivative functionals (IDDF) involving an arbitrary derivative order (for details see Section \ref{sec-iddf}). IDDFs are crucial elements in almost every modern bandwidth selection algorithm.

The remainder of the paper is organized as follows: in Section \ref{sec-band-selectors} we give an overview of the most popular and the most frequently used bandwidth selectors. In Section \ref{sec-prob-demo}, based on a simple example, we demonstrate the problem. In Section \ref{sec-fft-based-impl} we give details of a complete FFT-based algorithm for fast estimation of unconstrained bandwidth matrices. To make the presentation of our algorithm clear, we do it on the basis of one of the simplest bandwidth selection algorithm, namely least square cross validation (LSCV). In Section \ref{sec-iddf} we extend our results also for  the IDDFs. In Section \ref{sec-experim-results} we give results from some numerical experiments based on both synthetic and real data sets. In Section \ref{sec-conclusion} we conclude our paper. 

\section{Problem demonstration}\label{sec-prob-demo}
\begin{figure}[]
  \begin{minipage}[b]{0.5\textwidth}\begin{center}
   \includegraphics[scale=0.4]{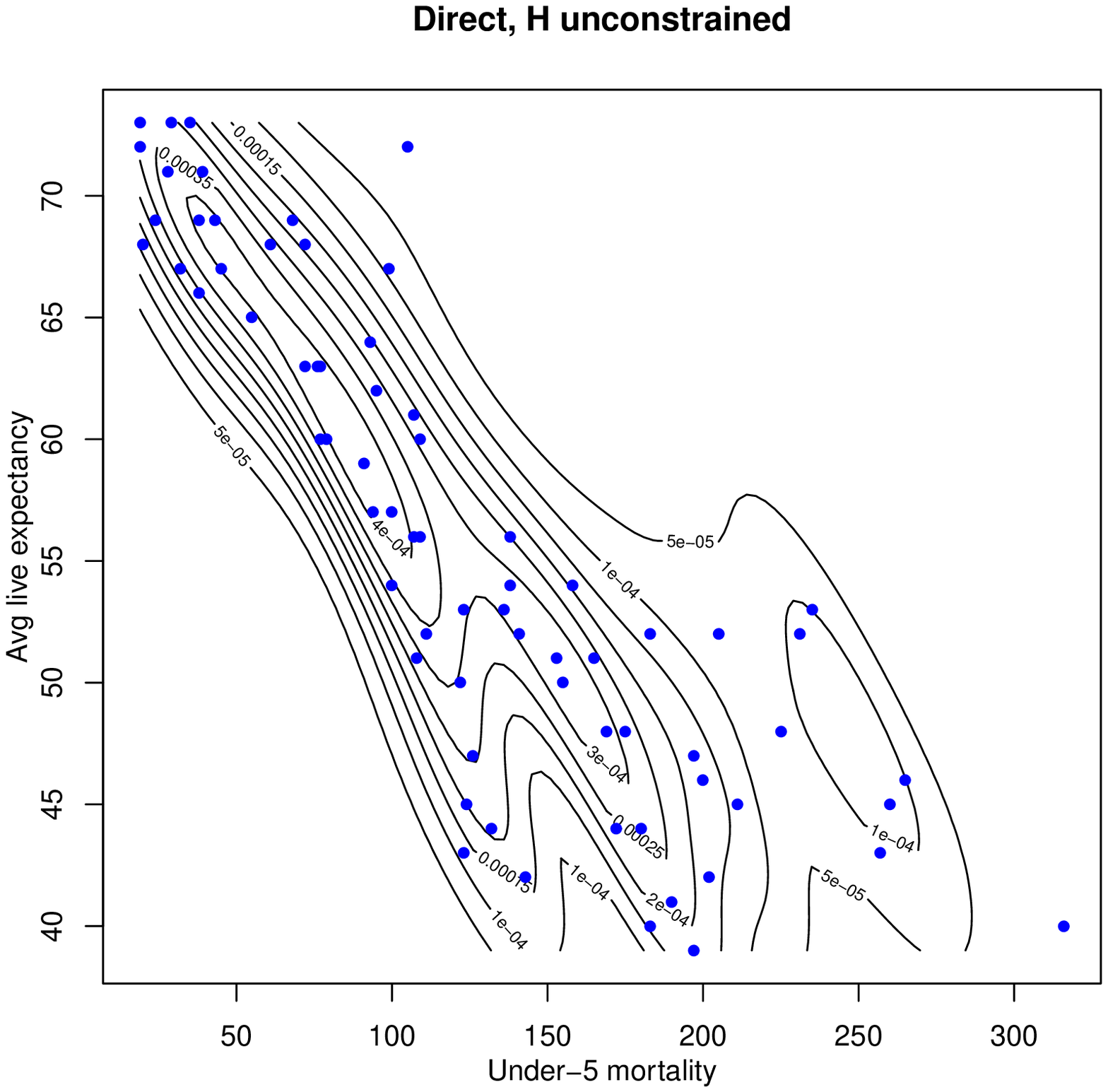} \\
   (a) \\  \vspace{0.5cm}
  \end{center}\end{minipage}
  \begin{minipage}[b]{0.5\textwidth}\begin{center}
   \includegraphics[scale=0.4]{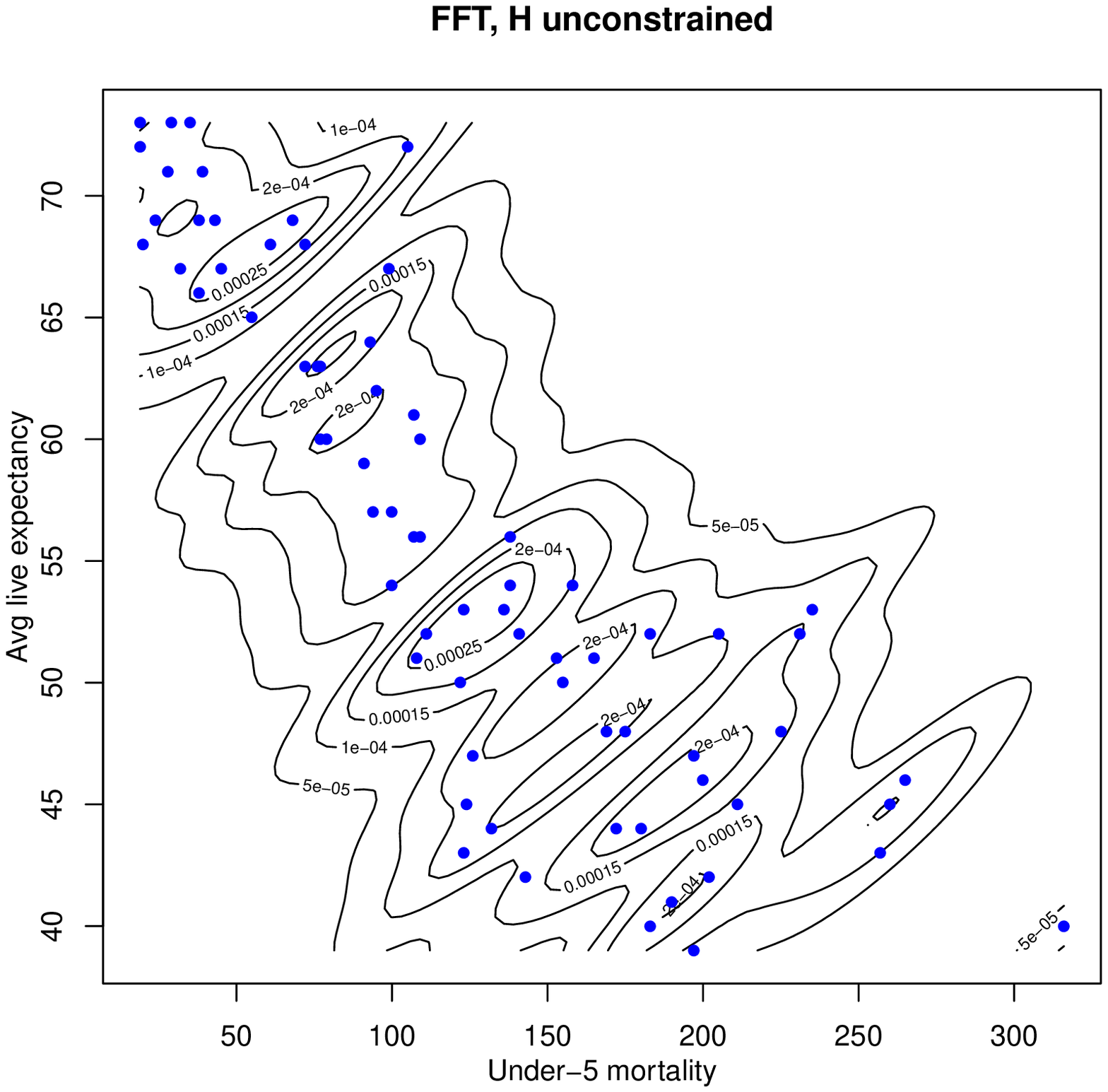} \\
   (b) \\  \vspace{0.5cm}
  \end{center}\end{minipage}
  \begin{minipage}[b]{0.5\textwidth}\begin{center}
   \includegraphics[scale=0.4]{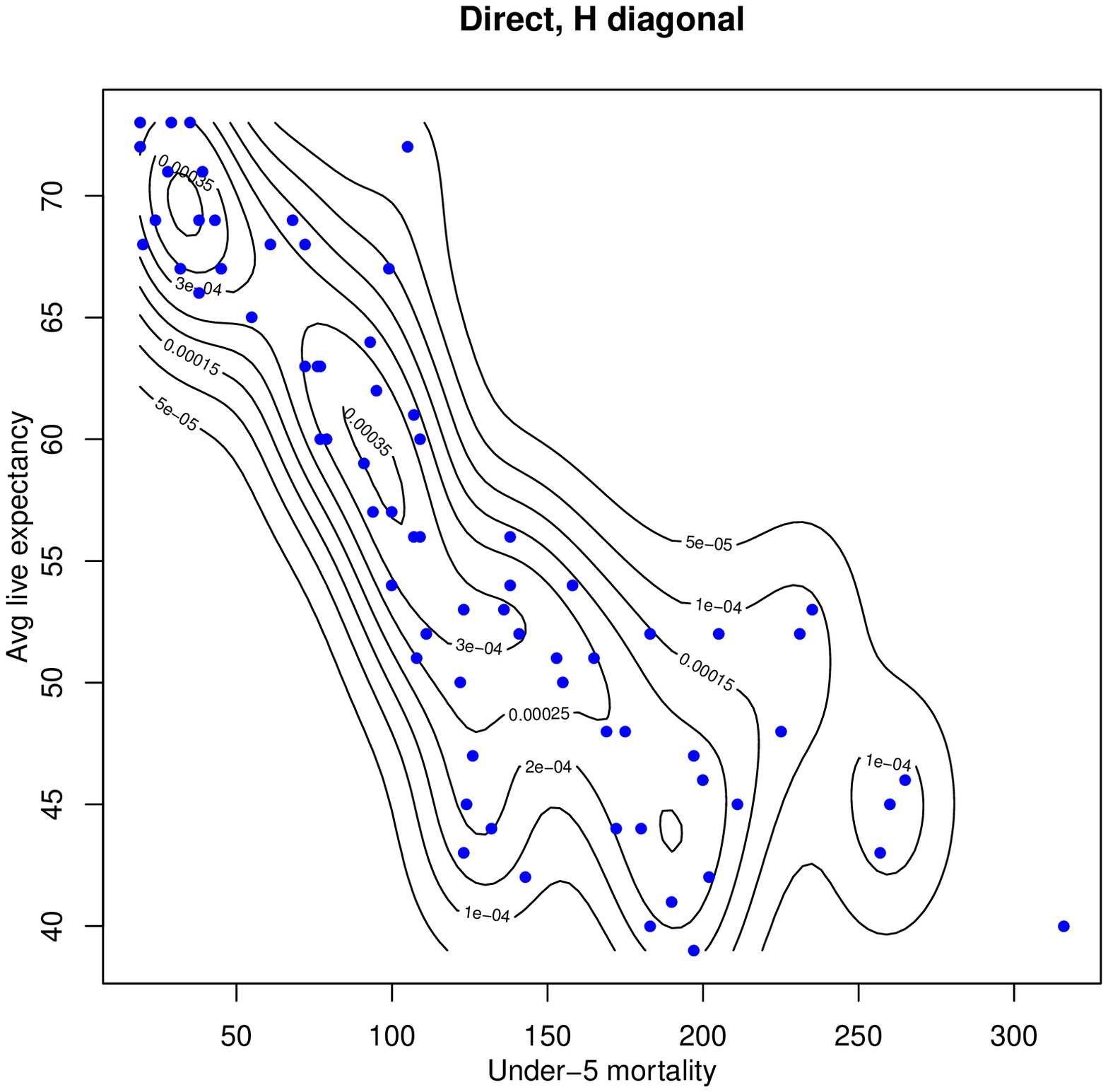} \\
   (c) \\  \vspace{0.0cm}
  \end{center}\end{minipage}
  \begin{minipage}[b]{0.5\textwidth}\begin{center}
   \includegraphics[scale=0.4]{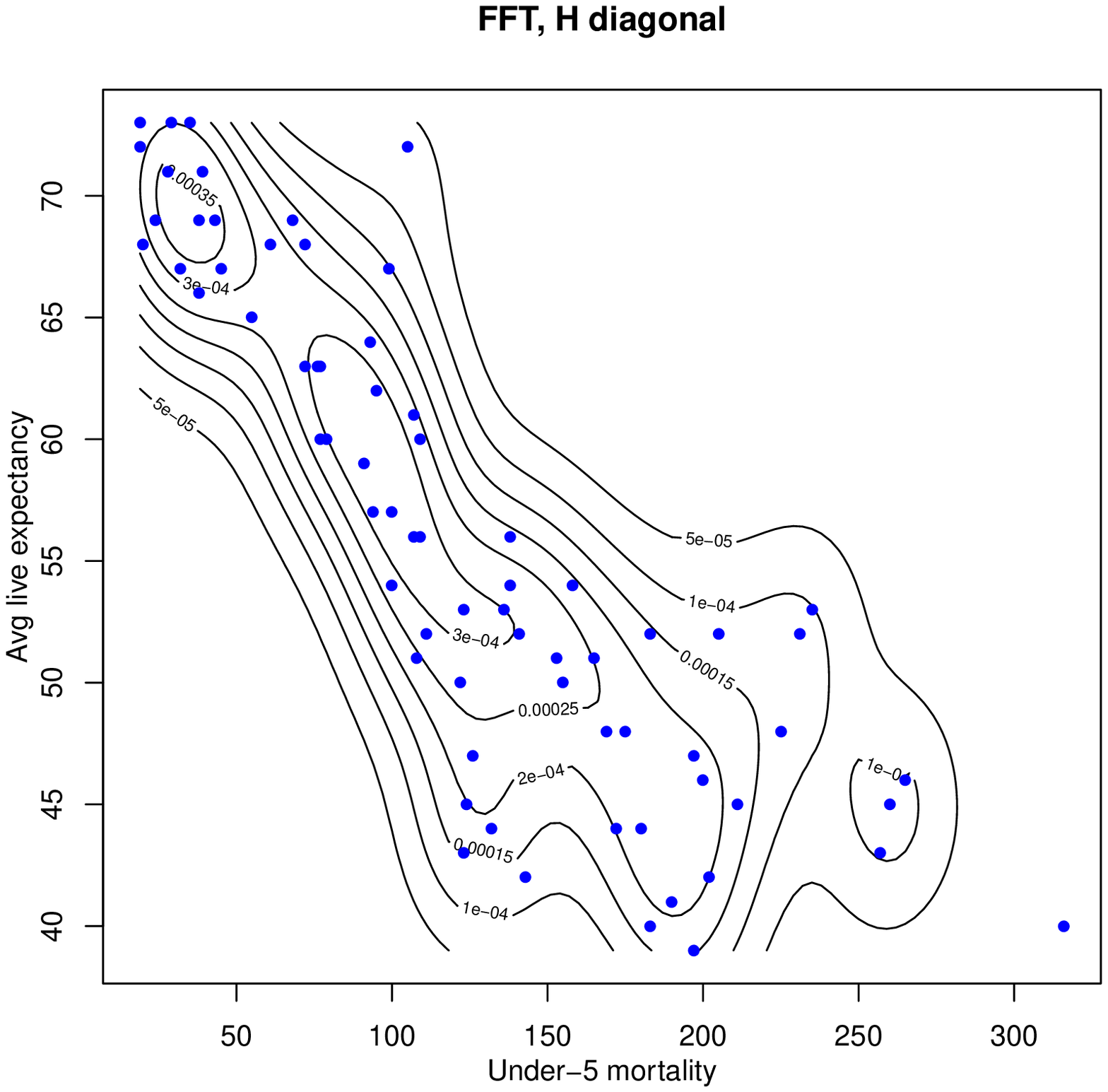} \\
   (d) \\  \vspace{0.0cm}
  \end{center}\end{minipage}
  \caption{Demonstration of behavior of Wand's original algorithm. (a) the reference density, (b) the behavior of Wand's original algorithm when the minimization of the objective function proceeds over $\boldsymbol{H} \in  \mathcal{F}$. The density is totally corrupted, (c) the reference density when the bandwidth was obtained by direct (i.e., non-FFT-based) implementation of Eq.~(\ref{eq-lscv-obj-fun}) when the minimization of the objective function now proceeds over $\boldsymbol{H} \in \mathcal{D}$, (d) the behavior of Wand's original algorithm when $\boldsymbol{H} \in \mathcal{D}$. Figures (c) and (d) are in fact almost identical.} 
  	\label{fig-wands-algo}
\end{figure}

As was mentioned in Section \ref{sec-intro}, Wand's algorithm does not support unconstrained bandwidth matrices, which considerably limits its practical usability. In this short demonstration we use a sample dataset \emph{Unicef} presented in more detail in Section \ref{sec-real-data}. In Fig.~\ref{fig-wands-algo}(a) we show the reference density when the bandwidth was obtained by direct (i.e., non-FFT-based) implementation of the LSCV algorithm, briefly presented in Section~ \ref{sec-band-selectors}. After numerical minimization of the resulting objective function we get the sought bandwidth. In Fig.~\ref{fig-wands-algo}(b) one can see the behavior of Wand's original algorithm (i.e., FFT-based) when the minimization of the objective function proceeds over $\boldsymbol{H} \in  \mathcal{F}$. The density is totally corrupted. In Fig.~\ref{fig-wands-algo}(c) we show the reference density when the bandwidth was obtained by direct (i.e., non-FFT-based) implementation of the LSCV algorithm when the minimization of the objective function now proceeds over $\boldsymbol{H} \in \mathcal{D}$. Finally, in Fig.~\ref{fig-wands-algo}(d) we show the behavior of Wand's original algorithm when $\boldsymbol{H} \in \mathcal{D}$. Figures  \ref{fig-wands-algo}(c) and  \ref{fig-wands-algo}(d) are practically identical, which confirms the fact that the original version of Wand's algorithm is adequate only for constrained bandwidth matrices. Some minor differences between Fig.~\ref{fig-wands-algo}(c)~and~\ref{fig-wands-algo}(d) are due to the binning of the original input data, but they are not of practical relevance. 

\begin{figure}[]
  \begin{minipage}[b]{0.5\textwidth}\begin{center}
   \includegraphics[scale=0.40]{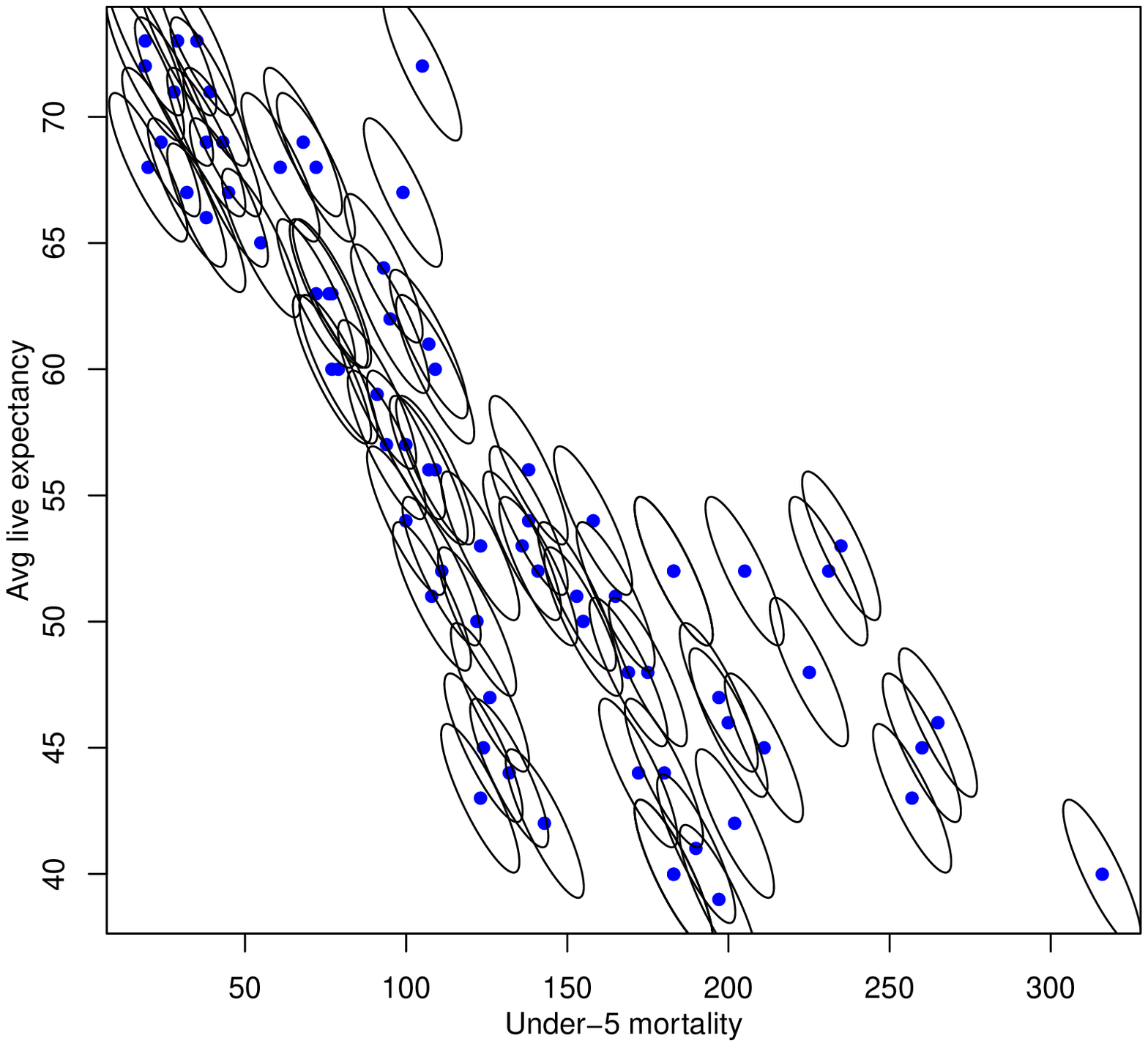} \\
   (a) \\  \vspace{0.0cm}
  \end{center}\end{minipage}
  \begin{minipage}[b]{0.5\textwidth}\begin{center}
   \includegraphics[scale=0.40]{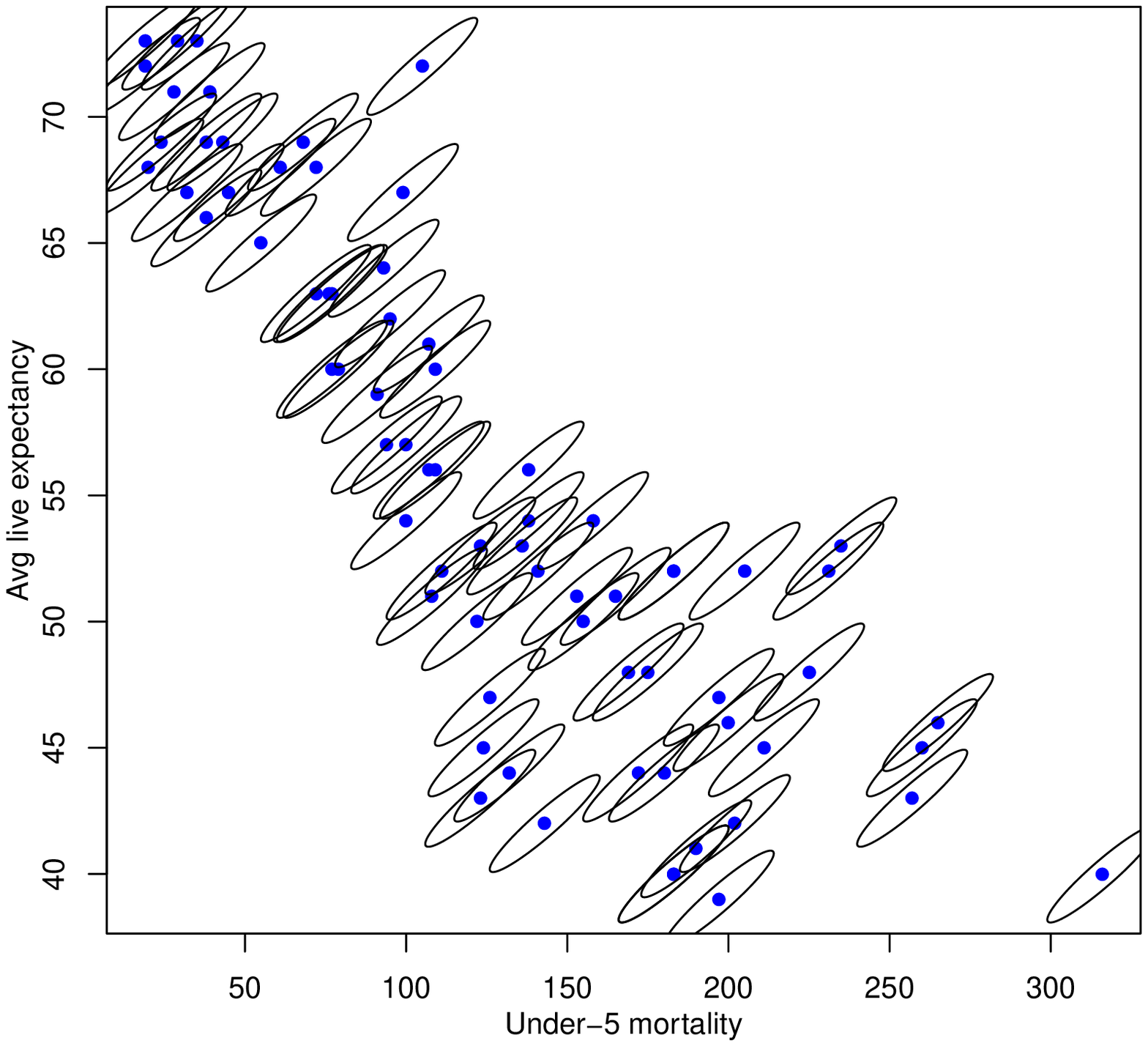} \\
   (b) \\  \vspace{0.0cm}
  \end{center}\end{minipage}
  \caption{Visualization of kernels $K_{\boldsymbol{H}}$ used for computing density $\hat{f}(\boldsymbol{x},\boldsymbol{H})$ for the sample dataset \emph{Unicef} (marked as small filled circles): (a) kernels generated by $\boldsymbol{H}_a$, (b) kernels generated by $\boldsymbol{H}_b$.} 
  	\label{fig-kernel-contours}
\end{figure}

The estimated bandwidth matrices used to plot densities shown in Figs.  \ref{fig-wands-algo}(a)--\ref{fig-wands-algo}(d) are as follows: 
\begin{align}\label{eq-Ha-Ha}
\boldsymbol{H}_a &=
\begin{bmatrix}
452.34 & -93.96 \\
-93.96 &  26.66 \\
\end{bmatrix},
\;\;\;
\boldsymbol{H}_b =
\begin{bmatrix}
896.20 & 94.98 \\
 94.98 & 11.37 \\
\end{bmatrix}, \nonumber \\
\boldsymbol{H}_c &=
\begin{bmatrix}
197.41 & 0.00 \\
 0.00 & 11.70 \\
\end{bmatrix},
\;\;\;
\boldsymbol{H}_d =
\begin{bmatrix}
242.42 & 0.00 \\
 0.00 & 11.97 \\
\end{bmatrix}.
\end{align}
It is easy to notice that in this particular example the off-diagonal entries in $\boldsymbol{H}_b$ are positive, while the `true' entries should be negative, as in $\boldsymbol{H}_a$. In the context of this example, this means that individual kernels $K_{\boldsymbol{H}}$ in Eq.~(\ref{eq-kde}) used for computing the density $\hat{f}(\boldsymbol{x},\boldsymbol{H})$ are (incorrectly) `rotated' about 90 degrees, as can be visualized in Fig.~\ref{fig-kernel-contours}. The kernels generated by $\boldsymbol{H}_a$ bandwidth follow correctly the north-west dataset orientation, while $\boldsymbol{H}_b$ bandwidth incorrectly generates north-east oriented kernels.

\section{Bandwidth selectors}\label{sec-band-selectors}

The accuracy of the kernel density estimators depend very strongly on the bandwidth. In the univariate case the bandwidth is a scalar entity which controls the amount of smoothing. In the multivariate case the bandwidth is a matrix which controls both the amount and the orientation of smoothing. This matrix can be defined on various levels of complexity. The simplest case is when a positive constant scalar multiplies the identity matrix, that is, $\boldsymbol{H} \in \mathcal{S}$ where $\mathcal{S}=\{h^2\boldsymbol{I_d}:h>0\}$. Another level of sophistication is $\boldsymbol{H} \in \mathcal{D}$. These two forms are often called \emph{constrained}. In the most general form the bandwidth is \emph{unconstrained}, that is, $\boldsymbol{H} \in  \mathcal{F}$. A very important problem before evaluating Eq.~(\ref{eq-kde}) is to find the optimal (under certain criteria) bandwidth and many original methods have been developed so far, most of them being automatic or data-driven bandwidth selectors. 

The problem of selecting the scalar bandwidth in univariate kernel density estimation is quite well understood. A number of methods exist that combine good theoretical properties with strong practical performance. See for example \cite{Jones:1996a}, \cite{Jones:1996b} and \cite{Wand-1995} where one can find a comprehensive history of these selectors. Many of these univariate selectors can be extended to the multivariate case in a relatively straightforward fashion if $\boldsymbol{H}$ is constrained (see \cite{Wand-1994b}, \cite{Sain:1994}). 
However, if $\boldsymbol{H}$ is unconstrained, such generalization is not so easy. 
Comprehensive analysis of unconstrained bandwidth selectors was made mainly in the following works: \cite{Duong:2004}, \cite{Duong:2005b}, \cite{Duong:2005a}, \cite{Chacon:2010}, \cite{Chacon:2011}, \cite{Chacon:2013} and \cite{Chacon:2015}. They provide references to all main bandwidth selectors, so here we do not reproduce them and we recommend the reader interested in details to consult these  references. 

Three major types of bandwidth selectors are: (a) methods which use very simple and easy to compute mathematical formulas; they were developed to cover a wide range of situations, but do not guarantee being close enough to the optimal (under certain criteria) bandwidth; they are often called the \emph{rules-of-thumb} (see \cite{Silverman:1986} and \cite{Scott:1992}), (b) methods based on \emph{cross-validation} (CV) ideas and more precise mathematical arguments (see \cite{Rudemo:1982} and \cite{Bowman:1984}). They require much more computational efforts, however, in reward for it, we get bandwidths which are more accurate for a wider class of density functions. Three classical variants of the CV methods are: \emph{least squares cross validation} (LSCV), sometimes called  \emph{unbiased cross validation} (UCV), see \cite{Scott:1987}, \emph{biased cross validation} (BCV), see \cite{Scott:1987}, and \emph{smoothed cross validation} (SCV), see \cite{Hall:1992}, (c) methods based on plugging in estimates of some unknown quantities that appear in formulas for the asymptotically optimal bandwidth. They are often called \emph{plug-in} (PI), see \cite{Park:1990} and \cite{Sheather:1991}. 

All CV-like selectors are based on estimating MISE or AMISE error criteria and then on  minimization of an objective function. In the case of the LSCV method such a function is defined in the following form:
\begin{align}\label{eq-lscv-obj-fun-def}
LSCV(\boldsymbol{H}) = 
\int_{\mathbb{R}^d} \hat{f}(\boldsymbol{x, \boldsymbol{H}})^2 d\boldsymbol{x}
-
2 n^{-1} \sum_{i=1}^{n}\hat{f}_{-i}(\boldsymbol{X}_i, \boldsymbol{H}),
\end{align}
where
\begin{align}\label{eq-leave-one-out}
\hat{f}_{-i}(\boldsymbol{x}, \boldsymbol{H})
=
(n-1)^{-1} \sum_{\substack{j=1\\j \neq i}}^{n} K_{\boldsymbol{H}}(\boldsymbol{x} - \boldsymbol{X}_j),
\end{align}
is the \emph{leave-one-out} estimator of $f$. Then, the LSCV objective function can be expressed as follows (for details, see \cite{Wand-1995}):
\begin{align}\label{eq-lscv-obj-fun}
LSCV(\boldsymbol{H}) 
&=
n^{-2} \sum_{i=1}^n \sum_{j=1}^n (K_{\boldsymbol{H}} * K_{\boldsymbol{H}})
(\boldsymbol{X}_i - \boldsymbol{X}_j) \nonumber \\
&- 
2 n^{-1} (n-1)^{-1} \sum_{i=1}^n \sum_{\substack{j=1\\j \neq i}}^n 
K_{\boldsymbol{H}}(\boldsymbol{X}_i - \boldsymbol{X}_j),
\end{align}
where $*$ denotes the convolution operator. For most practical implementations the normal kernels are used, i.e., $K=\Phi$, and then we obviously have $K_{\boldsymbol{H}} * K_{\boldsymbol{H}} = K_{\boldsymbol{2H}}$. The LSCV bandwidth matrix $\hat{\boldsymbol{H}}_{LSCV}$ is the minimizer of $LSCV(\boldsymbol{H})$, that is,
\begin{align}\label{eq-H-minimizer}
\hat{\boldsymbol{H}}_{LSCV} = 
\text{arg} \, \operatorname*{min}_{\boldsymbol{H} \in \mathcal{F}} LSCV(\boldsymbol{H}).
\end{align}
 
In this paper the FFT-based algorithm is presented in details on the basis of the LSCV method as it is very popular among practitioners, mainly due to its intuitive motivation. However, this variant of the CV selector has some serious drawbacks recalled by many authors, like many local minima in the objective function and high variability (in the sense that for different datasets from the same distribution, it will typically give considerably different answers). Extending our results also for other CV selectors, as well as for the PI selector is also possible, see  Section \ref{sec-iddf} for details.

\section{FFT-based algorithm}\label{sec-fft-based-impl}
A preliminary work on using FFT to univariate kernel density estimation defined in  (\ref{eq-kde-1D}) was that in \cite {Silverman:1982}. Based on this idea, in \cite{Wand-1994a} the author proposed a more universal method for both uni- and multivariate cases and also gave a note on a possibility of using the FFT algorithm for computation of kernel functional estimates as they are particularly important in bandwidth selection for KDE.  Based on this note and using results given in \cite{Gramacki:2016} and in \cite{Chacon:2015}, in this paper we present a fast method for optimal unconstrained bandwidth selection. Below we present a complete procedure for the LSCV bandwidth selector where the FFT-based approach is used.

From the viewpoint of the main subject of this paper, Eq.~(\ref{eq-lscv-obj-fun}) has to be rewritten in a slightly different form. Our goal is to remove the unwanted condition $j \neq i$ in the second double summation. For a sufficiently large $n$ (several tens in practical applications) it is safe to assume $n \approx n-1$. Under this assumption, we can write the objective function in the following form
\begin{align}\label{eq-lscv-obj-fun-new}
LSCV(\boldsymbol{H}) 
=
n^{-2} \sum_{i=1}^n \sum_{j=1}^n T_{\boldsymbol{H}}(\boldsymbol{X}_i - \boldsymbol{X}_j)
+ 
2 n^{-1}K_{\boldsymbol{H}}(\boldsymbol{0}),
\end{align}
where
\begin{align}\label{eq-T-u}
T_{\boldsymbol{H}}(u) &= (K_{\boldsymbol{H}} * K_{\boldsymbol{H}})(u) - 2 K_{\boldsymbol{H}}(u), \nonumber \\
K_{\boldsymbol{H}}(\boldsymbol{0}) &= (2 \pi) ^ {-d/2} |\boldsymbol{H}|^{-1/2}.
\end{align}
Now we are interested in fast computation of the following part of Eq.~(\ref{eq-lscv-obj-fun-new}):
\begin{align}\label{eq-Psi-before-bin}
\psi(\boldsymbol{H}) 
=
n^{-2} \sum_{i=1}^n \sum_{j=1}^n T_{\boldsymbol{H}}(\boldsymbol{X}_i - \boldsymbol{X}_j).
\end{align} 
Computational complexity of Eq.~(\ref{eq-Psi-before-bin}) is clearly $O(n^2)$. Thus, its fast and accurate computation plays a crucial role in bandwidth selection problem. 

It is also easy to notice that Eq.~(\ref{eq-Psi-before-bin}) is a zero-th order derivative functional of the general $r$-th order form given by
\begin{align}\label{eq-functional}
\psi_r(\boldsymbol{H}) = 
n^{-2} \sum_{i=1}^{n} \sum_{j=1}^{n} \mathsf{D}^{\otimes r}K_{\boldsymbol{H}}
(\boldsymbol{X}_i -\boldsymbol{X}_j).
\end{align}
Here $r$ is the derivative order (even number), $\mathsf{D}$ is the gradient operator and the notation with $\otimes$ symbol is explained in details in Section \ref{sec-iddf} where the above mentioned generalization is presented in details.

In the \textbf{first step} the multivariate \emph{linear binning} of the input random variables $\boldsymbol{X}_i$ is required. The binning is a kind of data discretization, as described in \cite[Section 3]{Wand-1994a} and can be computed using a fast $O(n)$ algorithm by extending the `integer division' idea of \cite{Fan:1994}. The binned approximation of Eq.~(\ref{eq-Psi-before-bin}) is 

\begin{align}\label{eq-Psi-after-bin}
\tilde{\psi}(\boldsymbol{H}) 
&=
n^{-2}
\sum_{i_1=1}^{M_1} \cdots \sum_{i_d=1}^{M _d}
\sum_{j_1=1}^{M_1} \cdots \sum_{j_d=1}^{M _d} 
T_{\boldsymbol{H}}(\boldsymbol{g}_i - \boldsymbol{g}_j) \boldsymbol{c}_i \boldsymbol{c}_j \nonumber \\
&=
\sum_{i_1=1}^{M_1} \cdots \sum_{i_d=1}^{M _d} \boldsymbol{c}_i
\left(
\sum_{j_1=1}^{M_1} \cdots \sum_{j_d=1}^{M _d} 
T_{\boldsymbol{H}}(\boldsymbol{g}_i - \boldsymbol{g}_j) \boldsymbol{c}_j
\right),
\end{align}
where $\boldsymbol{g}$ are equally spaced \emph{grid points} and $\boldsymbol{c}$ are \emph{grid counts}. Grid counts are obtained by assigning certain weights to the grid points, based on neighbouring observations. In other words, each grid point is accompanied by a corresponding grid count. 

The following notation is used (taken from \cite{Wand-1994a}): for $k=1,\ldots,d$, let $g_{k1} < \cdots < g_{kM_K}$ be an equally spaced grid in the $k$th coordinate directions such that $[g_{k1}, g_{kM_k}]$ contains the $k$th coordinate grid points. Here $M_k$ is a positive integer representing the \emph{grid size} in direction $k$. Let
\begin{align}
\boldsymbol{g_j} =
(g_{1j_1}, \ldots, g_{dj_d}), \;\;\; 1 \le j_k \le M_k, \;\;\; k=1,\ldots, d,
\end{align}
denote the grid point indexed by $\boldsymbol{j}=(j_1,\ldots,j_d)$ and the $k$th binwidth  be denoted by
\begin{align}\label{eq-delta-k}
\delta_k = \frac{g_{kM_k} - g_{k1}} {M_k - 1}.
\end{align}

In the \textbf{second step}, the summation inside the brackets in Eq.~(\ref{eq-Psi-after-bin}) is rewritten so that it takes a form of the convolution
\begin{align}\label{eq-Psi-after-bin-2}
\tilde{\psi}(\boldsymbol{H}) 
&=
n^{-2}
\sum_{i_1=1}^{M_1} \cdots \sum_{i_d=1}^{M _d} 
\boldsymbol{c}_i
\left(
\sum_{j_1= -(M_1 - 1)}^{M_1 - 1} \cdots \sum_{j_d= -(M_d - 1)}^{M_d -1} 
\boldsymbol{c}_{i-j} \boldsymbol{k}_j 
\right) \nonumber \\
&=
\sum_{i_1=1}^{M_1} \cdots \sum_{i_d=1}^{M _d} 
\boldsymbol{c}_i \;\;
(\boldsymbol{c \star k}),
\end{align}
where
\begin{align}\label{eq-kj}
\boldsymbol{k}_j 
= 
T_{\boldsymbol{H}} (\delta_1 j_1, \dots, \delta_d j_d). 
\end{align} 

In the \textbf{third step}, we compute the convolution between $\boldsymbol{c}_{i-j}$ and $\boldsymbol{k}_j$ using the FFT algorithm in only $O(M_1 \log M_1 \ldots M_d \log M_d)$ operations compared to the  $O(M_1^2 \ldots M_d^2 )$ operations required for direct computation of Eq.~(\ref{eq-Psi-after-bin}). To compute the convolution between $\boldsymbol{c}$ and $\boldsymbol{k}$ they must first be reshaped (\emph{zero-padded}) according to precise rules which are described in detail in \cite{Gramacki:2016}. Here, for simplicity, only two-dimensional variant is presented as extension to higher dimensions is straightforward. We have
\begin{align}\label{eq-K-2D-new}
\boldsymbol{k}_{zp} =
\begin{bmatrix}
\boldsymbol{k} & \boldsymbol{0} \\
\boldsymbol{0} & \boldsymbol{0} \\
\end{bmatrix}
=
\begin{bmatrix}
k_{-M_1,-M_2} & \cdots & k_{-M_1,0} & \cdots & k_{-M_1,M_2} & \\
\vdots        & \ddots & \vdots     & \ddots & \vdots       & \\ 
k_{0,-M_2}    & \cdots & k_{0,0}    & \cdots & k_{0,M_2}    & \boldsymbol{0} \\
\vdots        & \ddots & \vdots     & \ddots & \vdots       & \\ 
k_{M_1,-M_2}  & \cdots & k_{M_1,0}  & \cdots & k_{M_1,M_2}  & \cdots \\
              &        & \boldsymbol{0} &        & \vdots       & \boldsymbol{0} \\
\end{bmatrix},
\end{align}
and
\begin{align}\label{eq-C-2D-new}
\boldsymbol{c}_{zp} =
\begin{bmatrix}
\boldsymbol{0} & \boldsymbol{0} & \boldsymbol{0} \\
\boldsymbol{0} & \boldsymbol{c} & \boldsymbol{0} \\
\boldsymbol{0} & \boldsymbol{0} & \boldsymbol{0} \\
\end{bmatrix}
=
\begin{bmatrix}
\boldsymbol{0} & \vdots     & \boldsymbol{0} & \vdots      & \boldsymbol{0} \\ 
\cdots     & c_{1,1}    & \cdots     & c_{1,M_2}   & \cdots \\
\boldsymbol{0} & \vdots     & \ddots     & \vdots      & \boldsymbol{0} \\
\cdots     & c_{M_1,1}  & \cdots     & c_{M_1,M_2} & \cdots \\
\boldsymbol{0} & \vdots     & \boldsymbol{0} & \vdots      & \boldsymbol{0} \\            
\end{bmatrix},
\end{align}
where the entry $c_{1,1}$ in (\ref{eq-C-2D-new}) is placed in row $M_1$  and column $M_2$.
The sizes of the zero matrices are chosen so that after reshaping of $\boldsymbol{c}$ and $\boldsymbol{k}$, they both have the same dimension $P_1 \times P_2, \times, \ldots, \times P_d$ (highly composite integers; typically, a power of 2). $P_k$ ($k=1, \ldots, d$) are computed according to the following equation
\begin{align}\label{eq-Pk}
P_k = 2^{ \big{\lceil} {\log_2(3M_k -1) \big{\rceil}}}.
\end{align}

Now, to evaluate the summations inside the brackets in Eq.~(\ref{eq-Psi-after-bin-2}), we can apply the discrete convolution theorem, that is, we must do the following operations: 
\begin{align}\label{eq-disc-conv-theo}
\boldsymbol{C}&=\mathcal{F}(\boldsymbol{c}_{zp}), \;\;\; 
\boldsymbol{K}=\mathcal{F}(\boldsymbol{k}_{zp}), \;\;\;
\boldsymbol{S}=\boldsymbol{C}\boldsymbol{K}, \;\;\; 
\boldsymbol{s}=\mathcal{F}^{-1}(\boldsymbol{S}), 
\end{align}
where $\mathcal{F}$ stands for the Fourier transform and  $\mathcal{F}^{-1}$ is its inverse. The sought convolution $(\boldsymbol{c \star k})$ corresponds to a subset of $\boldsymbol{s}$ in Eq.~(\ref{eq-disc-conv-theo})  divided by the product of $P_1,P_2,\ldots,P_d$ (the so-called normalization), that is,
\begin{align}\label{eq-c_star_k}
(\boldsymbol{c \star k}) = 
\frac{1}{(P_1 \; P_2 \ldots P_d)}
\boldsymbol{s} [(2M_1-1) : (3M_1-2), \ldots ,(2M_d-1) : (3M_d-2)], 
\end{align}
where, for the two-dimensional case, $\boldsymbol{s}[a:b, c:d]$ means a subset of rows from $a$ to $b$ and a subset of columns from $c$ to $d$ of the matrix $\boldsymbol{s}$. 

In the \textbf{fourth step}, to complete the calculations of Eq.~(\ref{eq-Psi-after-bin-2}), the resulting $d$-dimensional array $(\boldsymbol{c \star k})$ needs to be multiplied by the corresponding grid counts $\boldsymbol{c}_i$ and summed to obtain $\tilde{\psi}(\boldsymbol{H})$, that is,
\begin{align}\label{eq-Psi-final}
\tilde{\psi}(\boldsymbol{H}) = n^{-2} \sum_{i} (\boldsymbol{c}_i 
\odot (\boldsymbol{c \star k})), 
\end{align}
where $ \odot$ means the element-wise multiplication. Finally, the sought $LSCV(\boldsymbol{H})$ in Eq.~(\ref{eq-lscv-obj-fun-new}) can be easily and effectively calculated.

In practical implementations, the sum limits $\{M_1, \ldots, M_d\}$ can be additionally shrunk to some smaller values $\{L_1, \ldots, L_d\}$, which significantly reduces the computational burden (see Section \ref{sec-speed-comparisions} for numerical results). In most cases, the kernel $K$ is the multivariate normal density function and, as such, an \emph{effective support} can be defined, i.e., the region outside which the values of $K$ are practically negligible. Now Eq.~(\ref{eq-Psi-after-bin-2}) can be rewritten as
\begin{align}\label{eq-Psi-after-bin-3}
\tilde{\psi}(\boldsymbol{H}) 
=
n^{-2}
\sum_{i_1=1}^{M_1} \cdots \sum_{i_d=1}^{M _d} 
\boldsymbol{c}_i
\left(
\sum_{j_1= -L_1}^{L_1} \cdots \sum_{j_d= -L_d}^{L_d} 
\boldsymbol{c}_{i-j} \boldsymbol{k}_j 
\right).
\end{align}
We propose to calculate $L_k$ using the following formula ($k=1, \ldots, d$):
\begin{align}\label{eq-L}
L_k
= 
\min 
\left( 
M_k-1, 
\Bigg{\lceil}
\frac{\tau \; \sqrt{ | \lambda | }}{\delta_k} 
\Bigg{\rceil}
\right),
\end{align} 
where $\lambda$ is the largest eigenvalue of $\boldsymbol{H}$ and $\delta_k$ is the mesh size from Eq.~(\ref{eq-delta-k}). After some empirical tests we found that $\tau$ can be set to around $3.7$ for a standard two-dimensional normal kernel. Such a value of $\tau$ guarantees that $\tilde{\psi}(\boldsymbol{H})$ calculated by either~(\ref{eq-Psi-after-bin-2}) or (\ref{eq-Psi-after-bin-3}) differs very little. Finally, we can calculate sizes $P_k$ of matrices (\ref{eq-K-2D-new}) and (\ref{eq-C-2D-new}) according to the following equation
\begin{align}\label{eq-Pk-2}
P_k = 2^{  \big{\lceil} (\log_2(M_k + 2 L_k - 1)  \big{\rceil} }.
\end{align}

\section{Notes on FFT-based algorithm for integrated density derivative functionals}\label{sec-iddf}

We will start this section by a short introduction to some notation which will be used in further discussion. This notation was introduced in \cite{Chacon:2010} and then used by those authors in consecutive papers, see the Reference. Let $f$~be a real $d$-variate function and its first derivative (gradient) vector is defined as  $\mathsf{D}f= \partial f / \partial \boldsymbol{x} = (\partial f / \partial x_1, \ldots , \partial f / \partial x_d)$ with $\boldsymbol{x}=(x_1,\ldots,x_d)^T$. Then the $r$-th derivative of $f$ is defined to be the \emph{vector} $\mathsf{D}^{\otimes r}f \in \mathbb{R}^{d^r}$. According to this notation $\mathsf{D}^{\otimes r}$ denotes the $r$-th Kronecker power of the operator $\mathsf{D}$, formally understood as the $r$-fold product $\mathsf{D} \otimes \cdots \otimes \mathsf{D}$. Naturally, $\mathsf{D}^{\otimes 0}f=f$, $\mathsf{D}^{\otimes 1}f=\mathsf{D}f$. Accordingly, all the second order partial derivatives can be organized into the Hessian matrix and the Hessian operator can be formally written as $\mathsf{H}=\mathsf{D}\mathsf{D}^T$ and of course  $\mathsf{H}f=\partial^2f/(\partial\boldsymbol{x}\partial \boldsymbol{x}^T)$ is the matrix of size $d \times d$.
Then, the \emph{equivalent vectorized} form is $\mathsf{D}^{\otimes 2}f=\mathsf{vec} \, \mathsf{H} \, f \in \mathbb{R}^{d^2}$, where $\mathsf{vec}$ denotes the operator which concatenates the columns of a matrix into a single vector, see \cite{Henderson:1979}. For $r \geq 3$ it is not clear how to organize the set containing all the $d^r$ partial derivatives of order $d$ into a matrix-like manner, so the above presented $\mathsf{D}^{\otimes r}$ notation seems to be very convenient, clear and useful.

Many (if not the most) modern bandwidth selection algorithms involve computing Eq.~(\ref{eq-functional}) for a given \emph{even} number $r$, which can vary, depending on a concrete algorithm. This is usually the most time and resource consuming part of these algorithms. Thus, in this chapter, we are concerned for detailed explanation on how the FFT-based algorithm can improve the computation of Eq.~(\ref{eq-functional}).

It is easy to notice that two computational problems occur here. The first one is how to calculate efficiently the $r$-th order partial derivatives of the kernel function $K_{\boldsymbol{H}}$. The second problem is how to efficiently calculate the double sums. Promising algebraic solutions of these two problems have been recently developed in \cite{Chacon:2015} where some efficient recursive algorithms were proposed. Our FFT-based solution can be seen as a useful extension of these developments. 

Below we expand the results from Section \ref{sec-fft-based-impl} where the LSCV bandwidth selector was analysed. This selector  involves Eq.~(\ref{eq-functional}) in its simplest form, that is for $r=0$ (zero-th order derivative), see Eq.~(\ref{eq-Psi-before-bin}). Hence, the FFT-based calculation of Eq.~(\ref{eq-Psi-before-bin}) is in fact a straightforward extension of our solution presented in \cite{Gramacki:2016}. 

However, extending our FFT-based solution for a more general case when $r > 0$ is not so easy (in most practical applications $r=2,4,6,8$). The problem comes from the fact that such $r$-th derivative is the set af all its partial derivatives of order~$r$. Here $\mathsf{D}^{\otimes r}K_{\boldsymbol{H}}$ denotes the vector containing all the $r$-th partial derivatives of $K_{\boldsymbol{H}}$ and its length is $d^r$. The vector-like arranging (instead of, for example, as an $r$-fold tensor array or as a multivariate matrix) is preferred here as it significantly simplifies multivariate analysis. However, the $d^r$ length of $\psi_r(\boldsymbol{H})$ quickly run into computational difficulties of the FFT-based algorithm. Fortunately, we can use excellent results presented in \cite[Chapter 6.3]{Chacon:2015} where the authors show how one can rewrite the three most often used algorithms for bandwidth selection (i.e., PI, UCV, SCV) in such a way that they (instead of direct usage of Eq.~(\ref{eq-functional})) utilize the V-statistics of the following general form 
\begin{align}\label{eq-V-statistics}
V_{mn} = \frac{1}{n^m} \sum_{i_1=1}^n \cdots \sum_{i_m=1}^n g(x_{i_1}, x_{i_2}, \dots, x_{i_m}),
\end{align}
where $g$ is a symmetric kernel function.  $V_{mn}$ is called a V-statistics of degree $m$. 

In the context of bandwidth selectors  the following V-statistics of degree~2 based on higher order derivatives of the Gaussian density function has the special meaning
\begin{align}\label{eq-V-statistics-degree-2}
V_{2n} = n^{-2} \sum_{i=1}^n \sum_{j=1}^n \eta_{r,s}(\boldsymbol{X}_i - \boldsymbol{X}_j; \boldsymbol{A}, \boldsymbol{B}, \boldsymbol{\Sigma}),
\end{align}
where $\eta_{r,s}$ is a \emph{scalar function} (and this fact is crucial here!) defined as
\begin{align}\label{eq-eta-r-s}
\eta_{r,s}(\boldsymbol{x}; \boldsymbol{A}, \boldsymbol{B}, \boldsymbol{\Sigma}) =
\{
(\mathsf{vec}^T \boldsymbol{A})^{\otimes r} 
\otimes 
(\mathsf{vec}^T \boldsymbol{B})^{\otimes s} 
\}
\mathsf{D}^{\otimes 2r + 2s}
\Phi_{\boldsymbol{\Sigma}}(\boldsymbol{x}).
\end{align}
Here $\boldsymbol{A}, \boldsymbol{B}$ are $d \times d$ symmetric matrices. Based on the above, we can define also
\begin{align}\label{eq-eta-r-0}
\eta_r(\boldsymbol{x}; \boldsymbol{\Sigma}) 
\equiv 
\eta_{r,0}(\boldsymbol{x}; \boldsymbol{I}_d, \boldsymbol{\Sigma}) 
= 
(\mathsf{vec}^T \boldsymbol{I}_d)^{\otimes r} \mathsf{D}^{\otimes 2r} \Phi_{\boldsymbol{\Sigma}}(\boldsymbol{x}).
\end{align}
Hence, what is required in the context of the paper's main subject is the FFT-based implementation of Eq.~(\ref{eq-V-statistics-degree-2}). The key for its efficient implementation is to develop a fast algorithm to compute the $\eta_{r,s}$ functions. The most time consuming part in the formula for $\eta_{r,s}$ is computation of the $r$-th derivative $\mathsf{D}^{\otimes 2r + 2s} \Phi_{\boldsymbol{\Sigma}}(\boldsymbol{x})$ of the multivariate Gaussian density function. Hopefully, in \cite{Chacon:2015} the authors give the complete and efficient algorithm for this task. Now, as $\eta_{r,s}$ is a scalar function, the FFT-based implementation is a straightforward replication of the four-step's procedure presented in Section \ref{sec-fft-based-impl} (see also the supplemental material for the up-to-date \textsf{R} source codes). 

The algorithm for computing  $\eta_{r,s}$ has been implemented in the \texttt{ks} \textsf{R} package \citep{ks} starting from version 1.10.0.  This package contains, among others, the function \texttt{Qr.cumulant\{ks\}} which is a very efficient implementation of the V-statistics of the following form
\begin{align}\label{eq-Qr}
Q_r(\boldsymbol{\Sigma}) 
=
n^{-2} \sum_{i=1}^n \sum_{j=1}^n \eta_{r}(\boldsymbol{X}_i - \boldsymbol{X}_j; \boldsymbol{\Sigma}).
\end{align} 
We use this function to make some numerical experiments. We compare the implementation of Eq.~(\ref{eq-Qr}) from the \texttt{ks} \textsf{R} package with our FFT-based implementation and it seems that such a comparison is a good idea to look at the advantages of the latter. The results are presented in Section \ref{sec-eta-rs-speed-comparisions}.

Coming back to the LSCV FFT-based algorithm presented in details in Section \ref{sec-fft-based-impl} we can rewrite appropriate equations using the $\eta_{r,s}$ functions. Thus,  Eq.~(\ref{eq-Psi-before-bin}) can be rewritten as
\begin{align}\label{eq-Psi-before-bin-eta-rs}
\psi_0(\boldsymbol{H}) 
&=
n^{-2} \sum_{i=1}^n \sum_{j=1}^n
T_{\boldsymbol{H}}(\boldsymbol{X}_i - \boldsymbol{X}_j) \nonumber \\ 
&=	
n^{-2} \sum_{i=1}^n \sum_{j=1}^n 
\Big\{
\eta_{0,0}(\boldsymbol{X}_i - \boldsymbol{X}_j; \boldsymbol{2H})  
-
2  \eta_{0,0}(\boldsymbol{X}_i - \boldsymbol{X}_j; \boldsymbol{H})
\Big\},
\end{align} 
and accordingly Eq.~(\ref{eq-kj}) becomes
\begin{align}\label{eq-kj-eta-rs}
\boldsymbol{k}_j 
= 
\eta_{0,0} (\delta_1 j_1, \dots, \delta_d j_d; \boldsymbol{2H})
-
2 \eta_{0,0} (\delta_1 j_1, \dots, \delta_d j_d; \boldsymbol{H}). 
\end{align} 
The LCSV criterion of Eq.~(\ref{eq-lscv-obj-fun}) can be rewritten in the most general case, supporting bandwidth selection for kernel density derivative estimation involving an arbitrary order $r$, as
\begin{align}\label{eq-lscv-obj-fun-eta_rs}
LSCV_r(\boldsymbol{H}) 
&=
(-1)^r \Big\{
n^{-2} \sum_{i=1}^n \sum_{j=1}^n \eta_{r,0}(\boldsymbol{X}_i - \boldsymbol{X}_j; \boldsymbol{2H}) \nonumber \\ 
&- 
2 n^{-1} (n-1)^{-1} \sum_{i=1}^n \sum_{\substack{j=1\\j \neq i}}^n 
\eta_{r,0}(\boldsymbol{X}_i - \boldsymbol{X}_j; \boldsymbol{H}) \Big\}.
\end{align}
For $r=0$ Eqs.~(\ref{eq-lscv-obj-fun}) and (\ref{eq-lscv-obj-fun-eta_rs}) are obviously equivalents. Finally, the LSCV criterion from Eq.~(\ref{eq-lscv-obj-fun-new}) can be rewritten in the same way
\begin{align}\label{eq-lscv-obj-fun-new-eta_rs}
LSCV_r(\boldsymbol{H}) 
&=
(-1)^r
\Big\{ 
n^{-2} \sum_{i=1}^n \sum_{j=1}^n
\Big\{
\eta_{r,0}(\boldsymbol{X}_i - \boldsymbol{X}_j; \boldsymbol{2H})  
-
2  \eta_{r,0}(\boldsymbol{X}_i - \boldsymbol{X}_j; \boldsymbol{H})
\Big\}
\nonumber \\ 
&+ 
2 n^{-1}K_{\boldsymbol{H}}(\boldsymbol{0})\Big\},
\end{align}
and accordingly, for $r=0$ Eqs.~(\ref{eq-lscv-obj-fun-new}) and (\ref{eq-lscv-obj-fun-new-eta_rs}) are also equivalents. Now, the FFT-based implementation of (\ref{eq-lscv-obj-fun-eta_rs}) and (\ref{eq-lscv-obj-fun-new-eta_rs}) is only a straightforward usage of the results from Section \ref{sec-fft-based-impl}. 

Last but not least, we have not implemented the complete procedures for PI and SCV bandwidth selectors (covering both the case of kernel density estimation and also kernel density derivative estimation) as this should go beyond the scope of the paper. But we can surely expect that our FFT-based solution should improve their performance as well. The mathematical formulas for PI and SCV selectors shown in Section 6.3 of \cite{Chacon:2015} involve double sums with $\eta_{r,2}$ and $\eta_{r,0}$ functions, so it is obvious that our solution can be easily adopted there, and in consequence, the selectors should work faster then their non-FFT counterparts. 

\section{Experimental results}\label{sec-experim-results}
This section is divided into four parts. The first part reports a simulation study based on synthetic data (two-dimensional mixtures of normal densities). The advantage of using such target densities is that we can compute exact Integrated Squared Errors (ISE) 
\begin{align}\label{eq-ISE}
\text{ISE} \hat{f}(H) = 
\int_{\mathbb{R}^d} \left( \hat{f}(\boldsymbol{x}, \boldsymbol{H}) - f(\boldsymbol{x}) \right)
^2 d\boldsymbol{x}
\end{align} 
between the resulting kernel density estimates and the target densities. It was proven that the ISE of any normal mixture density has an explicit form, see for example \cite{Duong:2004}. 

The second part reports a simulation study based on two real datasets. Here, the most handy way to compare the results is to use contour plots. We also moved away from the general multivariate case to the bivariate case as the results and the target densities can be easily visualized on two-dimensional plots. 

The third part reports speed results when we compare computational times needed for estimation of the optimal bandwidth matrices for both FFT-based and non-FFT-based (direct) algorithms. Also, usability of reducing $M_k$ into $L_k$ is analyzed (see Eqs.~(\ref{eq-Psi-after-bin-3}) and (\ref{eq-L})).

Finally, the fourth part reports speed results when we compare computational times needed for the V-statistics computation defined in Eq.~(\ref{eq-Qr}). We compare results returned by the  \texttt{Qr.cumulant\{ks\}} \textsf{R} function and our FFT-based implementation.

All the calculations were conducted in  the \textsf{R} environment. Minimization of the objective function $LSCV(\boldsymbol{H})$ was carried out using the \texttt{optim\{stats\}} \textsf{R} function. The \texttt{Nelder-Mead} method was used with default scaling parameters, that is the reflection factor $\alpha=1.0$, the contraction factor $\beta=0.5$ and the expansion factor $\gamma=2.0$. This method was chosen as it 
is robust and works reasonably well for nondifferentiable functions. A disadvantage of the method is that it is relatively slow. Some numerical-like problems are also reported. 

\subsection{Synthetic data}\label{sec-synth-data}
\begin{figure}
	\centering
	\includegraphics[width=13.5cm]{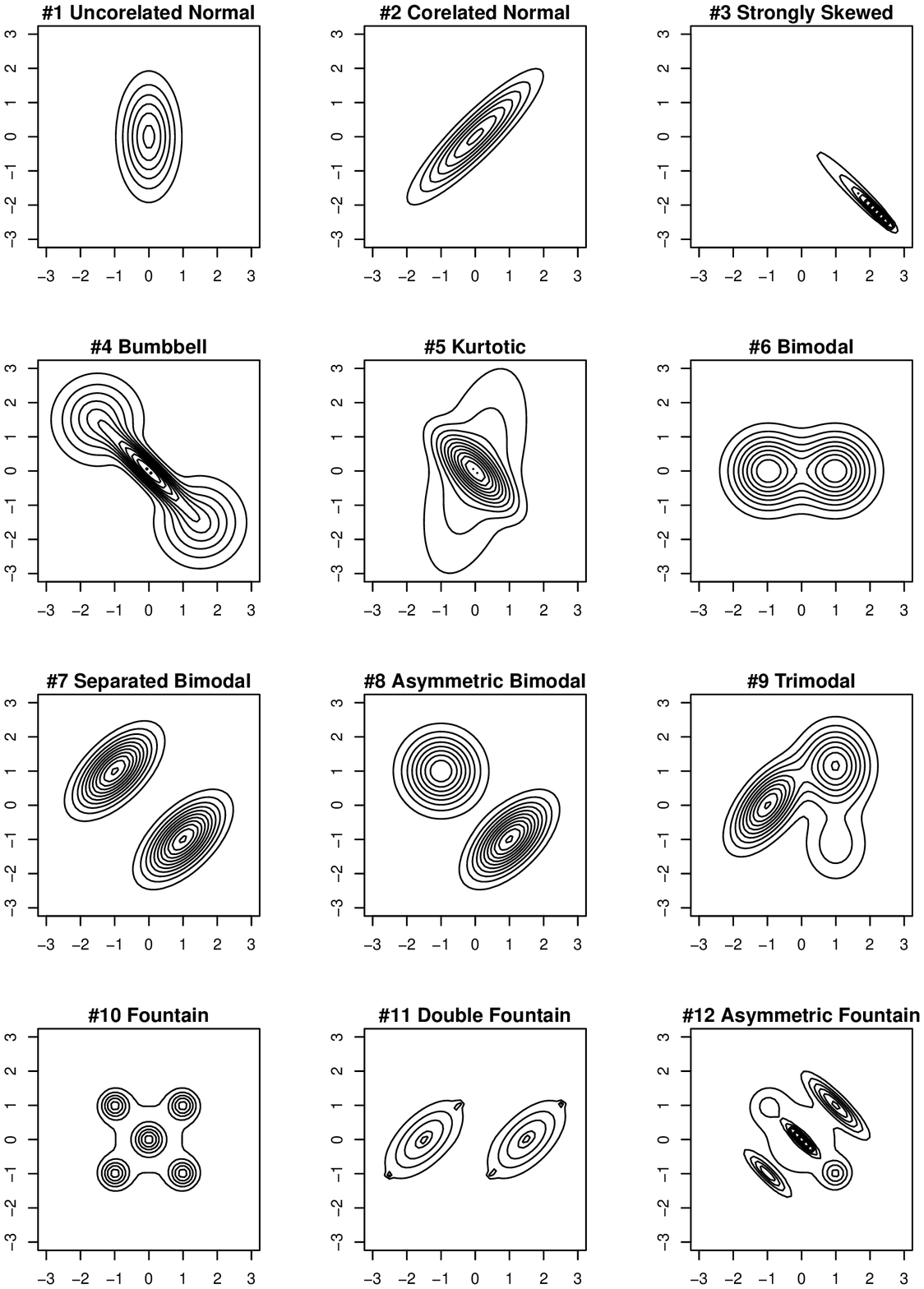}
	\caption{Contour plots for 12 target densities (normal mixtures).} 
	\label{fig-chacon-target-densities}
\end{figure}

The target densities which are analyzed were taken from \cite{Chacon:2009} as they cover a very wide range of density shapes. We preserve their original names and numbering. The shapes are shown in Fig. \ref{fig-chacon-target-densities}. 

We took sample sizes $n=\{128, 256, 1024\}$ and grid sizes (for simplicity equal in each direction) $M_1=M_2 =\{20, 25, 30, 35, 40, 45, 50, 150\}$. For each combination of the sample size and the grid size we computed the ISE error and these computations were repeated 50 times. In each repetition a different random sample was drawn from the target density. Then classical boxplots were drawn. We did not make separate simulations for $M_k$ and $L_k$ (see Eq.~(\ref{eq-Psi-after-bin-3}) and (\ref{eq-L})) as the results are practically the same for $\tau=3.7$.

\begin{figure}
	\centering
	\includegraphics[width=13.5cm]{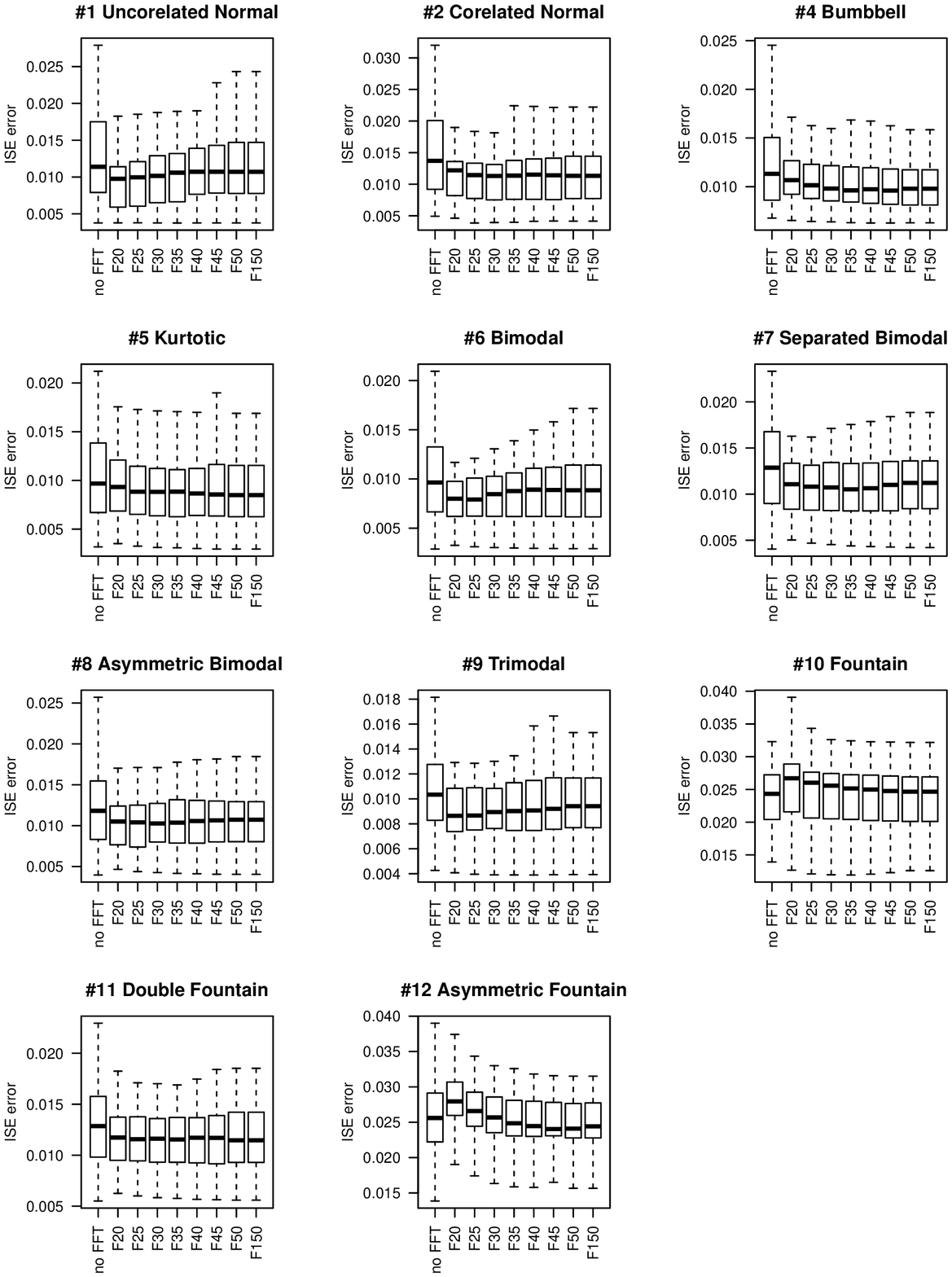}
	\caption{Boxplots of the ISE errors for the sample size $n=128$. `no FFT' is the reference boxplot calculated without the FFT, using the direct formula (\ref{eq-lscv-obj-fun-new}). Fxx means the boxplot where gridsize xx was use.} 
	\label{fig-boxplots-128}
\end{figure}

\begin{figure}
	\centering
	\includegraphics[width=13.5cm]{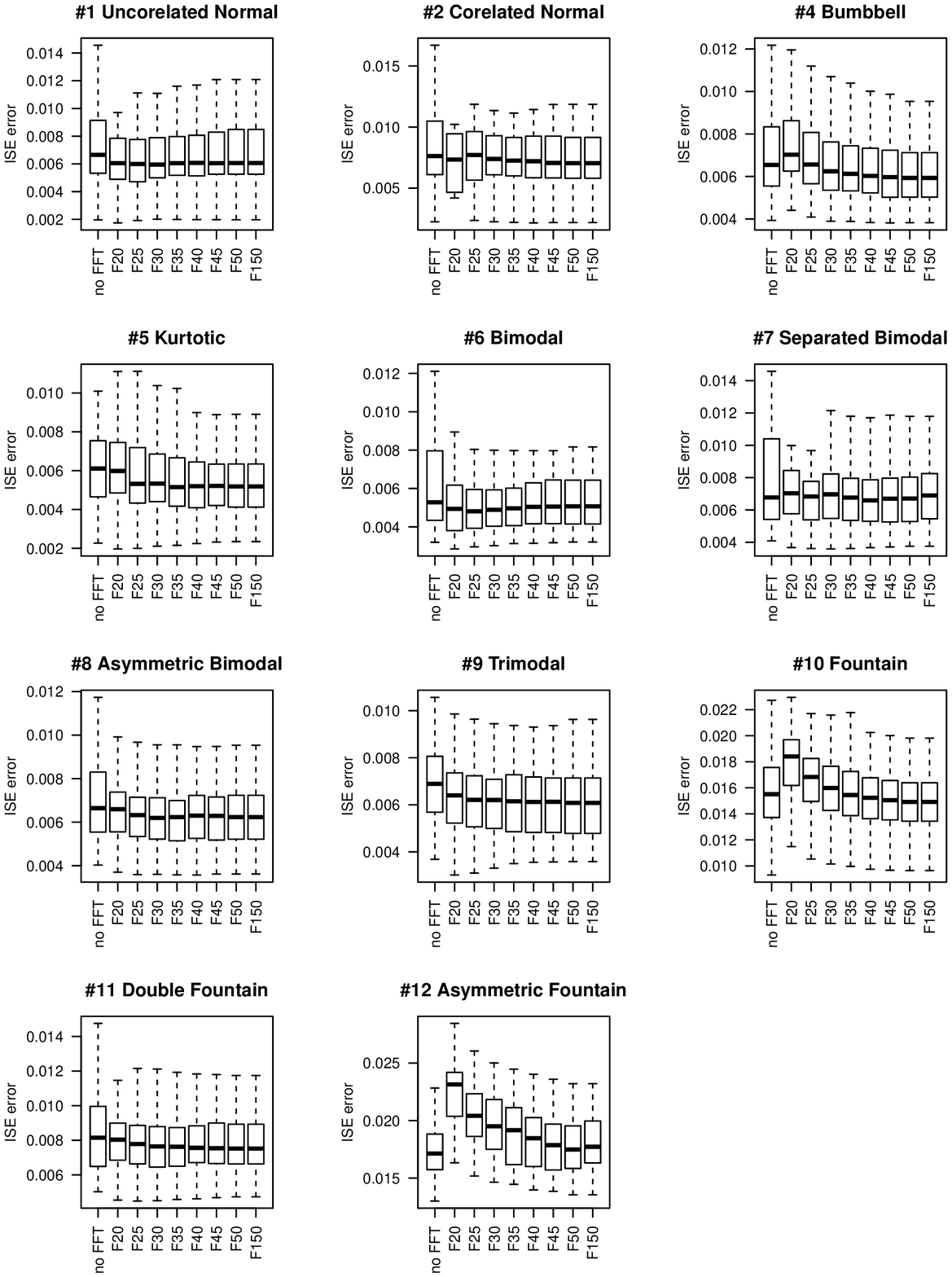}
	\caption{Boxplots of the ISE errors for the sample size $n=256$. `no FFT' is the reference boxplot calculated without the FFT, using direct the formula (\ref{eq-lscv-obj-fun-new}). Fxx means the boxplot where gridsize xx was use.  } 
	\label{fig-boxplots-256}
\end{figure}

Our goal was to check two things: (a) if, in general,  the FFT-based algorithm gives correct results (compared with a reference textual implementation based on Eq.~(\ref{eq-lscv-obj-fun-new})), and (b) how the binning operation may influence the final results. In Figs. \ref{fig-boxplots-128} and \ref{fig-boxplots-256} we present results for sample sizes $n=128$ and $n=256$, respectively. Looking at the boxplots we can see that the FFT-based solution is absolutely comparable to the direct solution. The ISE errors differ slightly, but from a practical point of view the fluctuations can be neglected.
\begin{figure}
	\centering
	\includegraphics[width=13.5cm]{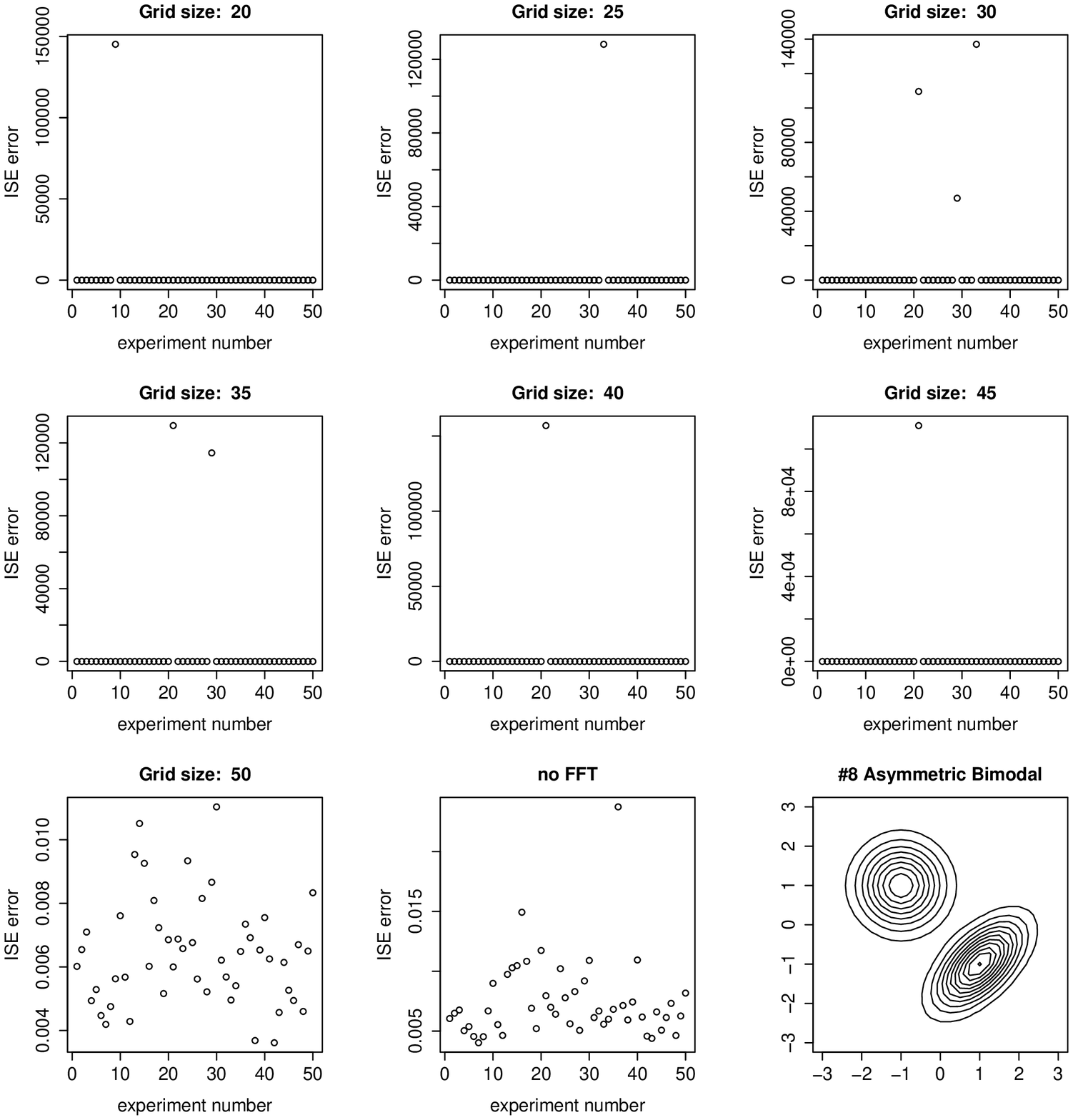}
	\caption{ISE errors for Model 8 and for each of 50 experiment repetitions. The sample size is $n=256$.} 
	\label{fig-ise-errors-model8}
\end{figure}

However, during practical experiments, problems of numerical nature were observed. First, we shall describe the problem, and next we shall try to give a plausible explanation. While preparing Figs. \ref{fig-boxplots-128} and \ref{fig-boxplots-256} only `good' results were used while `bad' results were discarded. The `bad' are these for which the derived ISE errors turned out to be extremely large, like many thousands or more.  In Fig.~\ref{fig-ise-errors-model8} we show every ISE error for Model 8 (Asymmetric Bimodal) and for each of the 50 experiment replications. As can be easily noticed, only direct $LSCV(\boldsymbol{H})$ computation (based on Eq.~(\ref{eq-lscv-obj-fun-new}), labeled `no FFT') and computations for the grid size equal to 50 yield all ISE errors on an acceptable level. Unfortunately, after binning the data, a number of failed optimizations occur, especially for smaller grids. In Table \ref{tab-ise-errors} we give exact numbers of optimization failures (that is, where we get excessive ISE errors) for some selected sample sizes, grid sizes and model numbers (see  Fig.~\ref{fig-chacon-target-densities}). The criterion for classifying a particular ISE error as excessive was $\text{ISE} > 1$. We can see (as expected) a pattern here that larger $n$, generates larger number of optimization failures. Another pattern is that increasing the grid size decreases the number of optimization failures. Moreover, some models are stiffer than others and it seems, e.g., that Model 3 is the most fragile and, in fact, FFT-based approach is unacceptable here. The problem with this particular model is caused by a specific data concentration, where most of the probability mass is concentrated on a very small area. Accordingly, in this case a denser griding is required.
\begin{table}[!ht]
\centering
\caption{Number of abnormally large ISE errors for particular models from Fig. \ref{fig-chacon-target-densities}. The criteria for classifying a particular ISE error as excessive was $\text{ISE} > 1$. }
\label{tab-ise-errors}
\small
\begin{tabular}{ccccccccccccc}
\hline
\multicolumn{13}{c}{$n=128$} \\
\hline
grid size & \#1 & \#2 & \#3 & \#4 & \#5 & \#6 & \#7 & \#8 & \#9 & \#10 & \#11 & \#12 \\
\hline
20        &  0  &  4  & 49  &  0  &  0  &  0  &  0  &  0  &  0  &  1   &  1   &  3 \\
30        &  0  &  0  & 21  &  0  &  0  &  0  &  0  &  0  &  0  &  0   &  0   &  2 \\
40        &  0  &  0  &  5  &  0  &  0  &  0  &  1  &  0  &  0  &  0   &  1   &  2 \\
50        &  0  &  0  &  1  &  0  &  0  &  0  &  0  &  0  &  0  &  0   &  0   &  1 \\
150       &  0  &  0  &  0  &  0  &  0  &  0  &  0  &  0  &  0  &  0   &  0   &  0 \\
\hline
\multicolumn{13}{c}{$n=256$} \\
\hline
grid size & \#1 & \#2 & \#3 & \#4 & \#5 & \#6 & \#7 & \#8 & \#9 & \#10 & \#11 & \#12 \\
\hline
20        &  0  & 44  & 50  &  0  &  6  &  0  &  3  &  1  &  0  &  5   &  0   & 10 \\
30        &  0  &  0  & 48  &  0  &  0  &  0  &  6  &  3  &  0  &  0   &  0   &  8 \\
40        &  0  &  0  & 36  &  0  &  0  &  0  &  5  &  1  &  0  &  0   &  0   &  5 \\
50        &  0  &  0  & 15  &  0  &  0  &  0  &  4  &  0  &  0  &  0   &  0   &  3 \\
150       &  0  &  0  &  1  &  0  &  0  &  0  &  0  &  0  &  0  &  0   &  0   &  0 \\
\hline
\multicolumn{13}{c}{$n=1024$} \\
\hline
grid size & \#1 & \#2 & \#3 & \#4 & \#5 & \#6 & \#7 & \#8 & \#9 & \#10 & \#11 & \#12 \\
\hline
20        &  0  & 50  & 50  & 48  & 50  &  0  & 29  &  8  &  0  &  50  &  19  & 50 \\
30        &  0  & 47  & 50  &  4  & 14  &  0  & 22  &  8  &  0  &  25  &   2  & 48 \\
40        &  0  &  0  & 50  &  0  &  1  &  0  & 38  & 11  &  0  &   2  &   0  & 36 \\
50        &  0  &  0  & 50  &  0  &  0  &  0  & 21  &  9  &  0  &   0  &   0  & 30 \\
150       &  0  &  0  &  4  &  0  &  0  &  0  &  1  &  0  &  0  &   0  &   1  &  2 \\
\hline
\end{tabular}
\end{table}

An obvious workaround of the above mentioned numerical problems can be increasing the grid size. Some suggestions about selection of grid sizes can be found in \cite{Gonzalez:1996} and are similar to our results.  According to the results given in Table \ref{tab-ise-errors} we can say that grid sizes of about $150 \times 150$ or more should be adequate in most practical applications.

What is also important, direct  $LSCV(\boldsymbol{H})$ minimization (that is without the FFT-based approach) is much more robust in the sense that there are no optimization problems as shown above. The explanation for this phenomena is that binning the data makes them highly discretized, even if there are no repeated values. This may result in a nondifferentiable objective function $LSCV(\boldsymbol{H})$ which is much more difficult for optimization algorithms, causing problems with finding a global minimum. Hence, more research is necessary to develop some new or improve the existing algorithms which will be more robust in the area of bandwidth estimation.  

\subsection{Real data}\label{sec-real-data}
In this section we analyze two real datasets. The first one is the well-known \emph{Old Faithful Geyser Data} as investigated in \cite{Azzalini:1990} (and many others). It consists of pairs of waiting times between eruptions and the durations of the eruptions for the Old Faithful geyser in Yellowstone National Park, Wyoming, USA. A data frame consists of 272 observations on 2 variables. The second dataset is the \emph{Unicef} one available in the \texttt{ks} \textsf{R} package \citep{ks}. This data set contains the numbers of deaths of children under 5 years per 1000 live births and the average life expectancy (in years) at birth for 73 countries with the GNI (Gross National Income) less than 1000 US dollars per annum per capita. A data frame consists of 73 observations on 2 variables. Each observation corresponds to a country. 

Here, the ISE criterion does not have a closed form (as opposed to any normal mixture densities used in Section \ref{sec-synth-data}), so the only sensible way to evaluate our FFT-based solution is to use contour plots. Before processing, all duplicates were discarded as all cross-validation methods are not well-behaved in this case. When there are duplicate observations, the procedure will tend to choose too small bandwidths.  
We did not make separate simulations for $M_k$ and $L_k$ (see Eqs.~(\ref{eq-Psi-after-bin-3}) and (\ref{eq-L})) as the results are practically the same for $\tau=3.7$.

\begin{figure}[]
  \begin{minipage}[b]{0.5\textwidth}\begin{center}
   \includegraphics[scale=0.31]{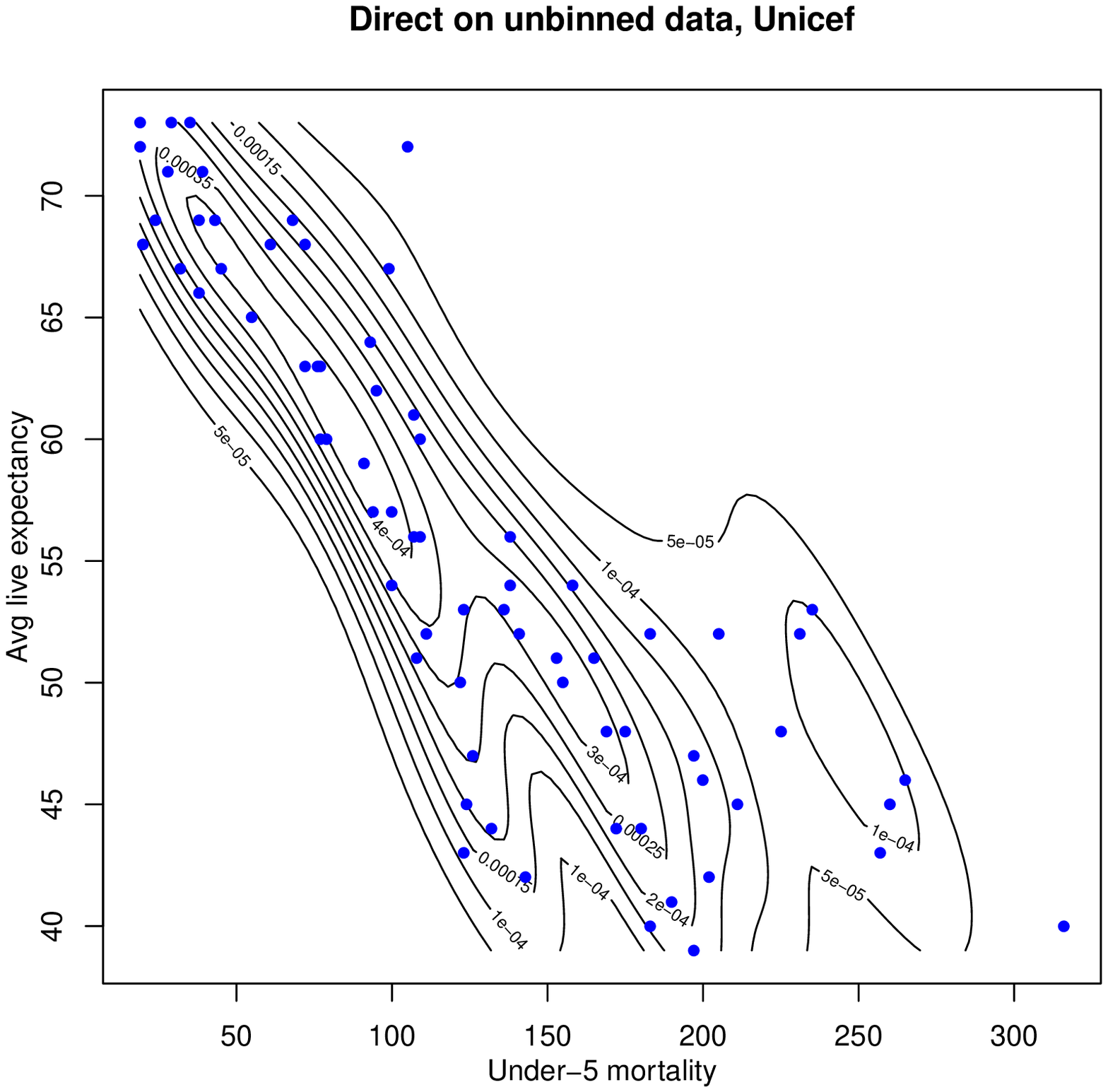} \\
   (a) \\  \vspace{0.2cm}
  \end{center}\end{minipage}
  \begin{minipage}[b]{0.5\textwidth}\begin{center}
   \includegraphics[scale=0.31]{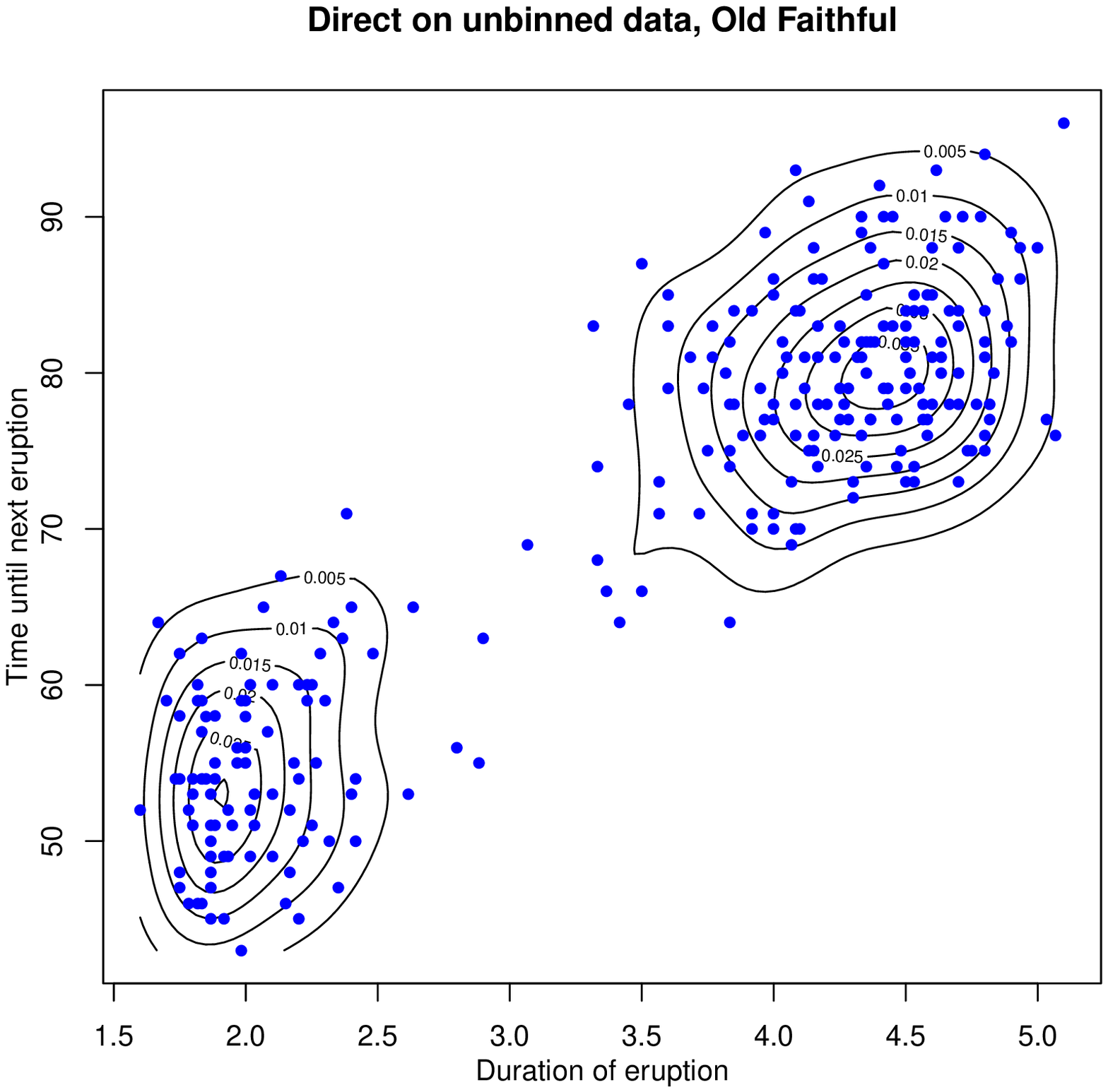} \\
   (b) \\  \vspace{0.2cm}
  \end{center}\end{minipage}
  \begin{minipage}[b]{0.5\textwidth}\begin{center}
   \includegraphics[scale=0.31]{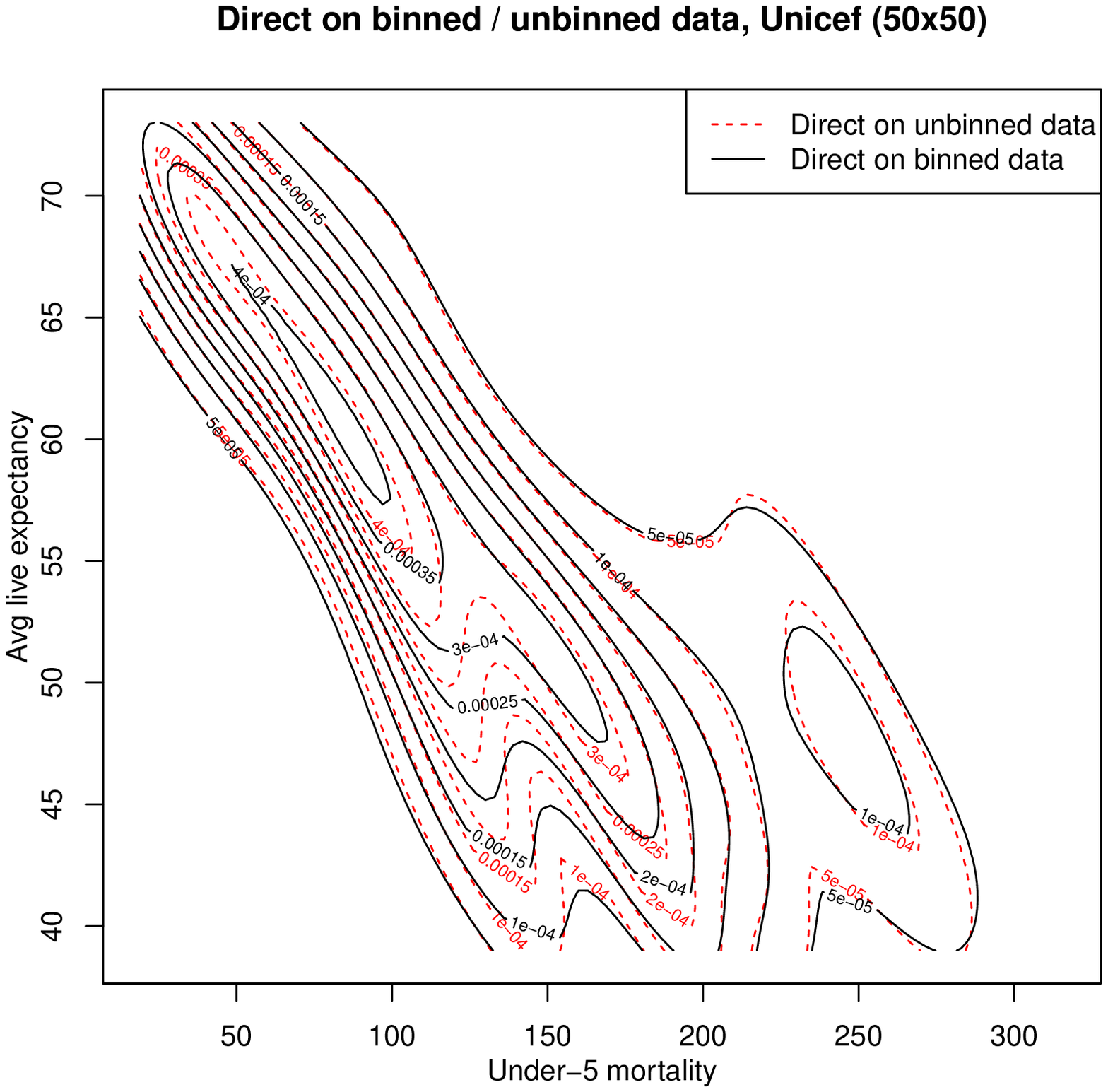} \\
   (c) \\  \vspace{0.2cm}
  \end{center}\end{minipage}
  \begin{minipage}[b]{0.5\textwidth}\begin{center}
   \includegraphics[scale=0.31]{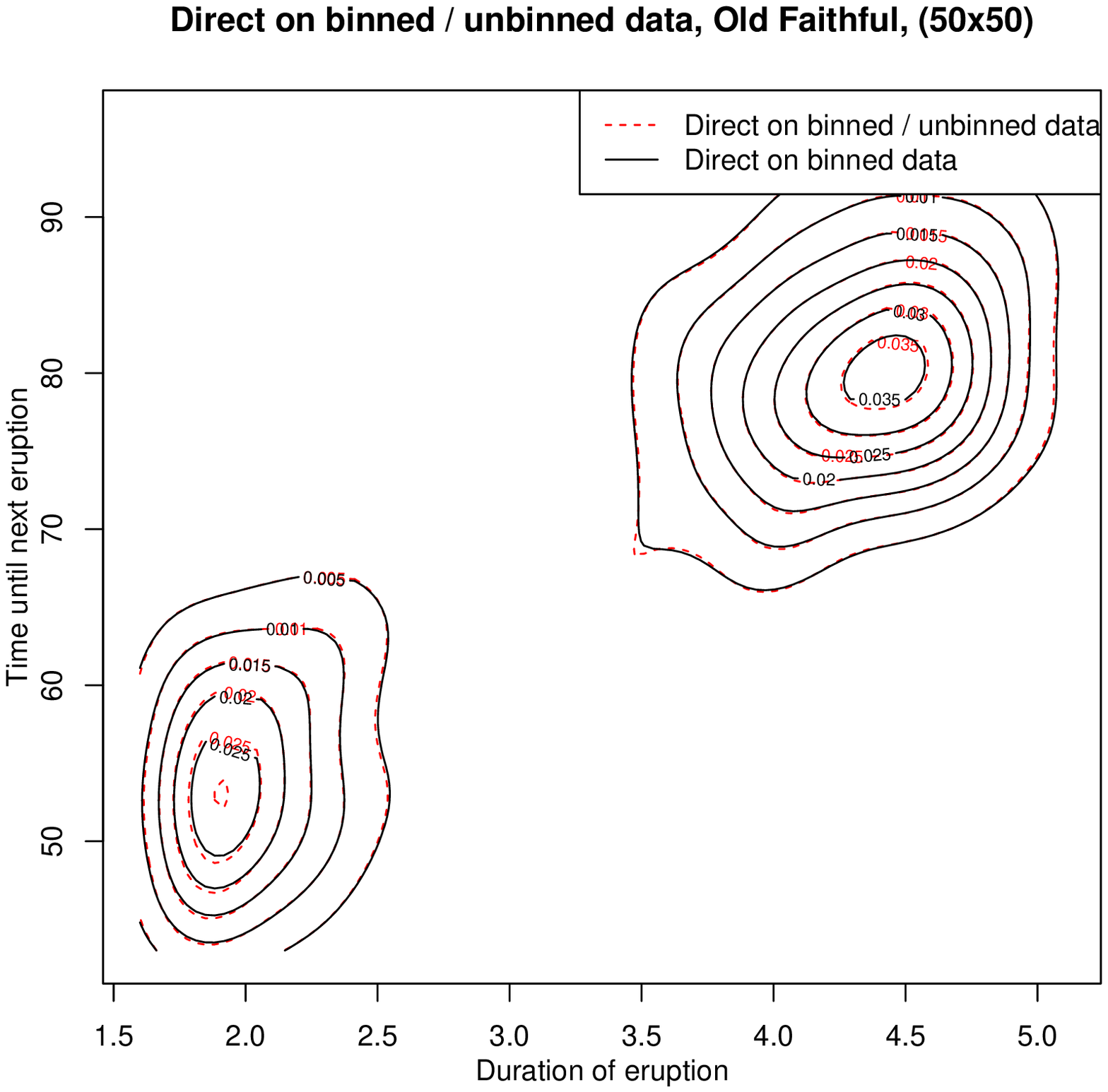} \\
   (d) \\  \vspace{0.2cm}
  \end{center}\end{minipage}
  \begin{minipage}[b]{0.5\textwidth}\begin{center}
   \includegraphics[scale=0.31]{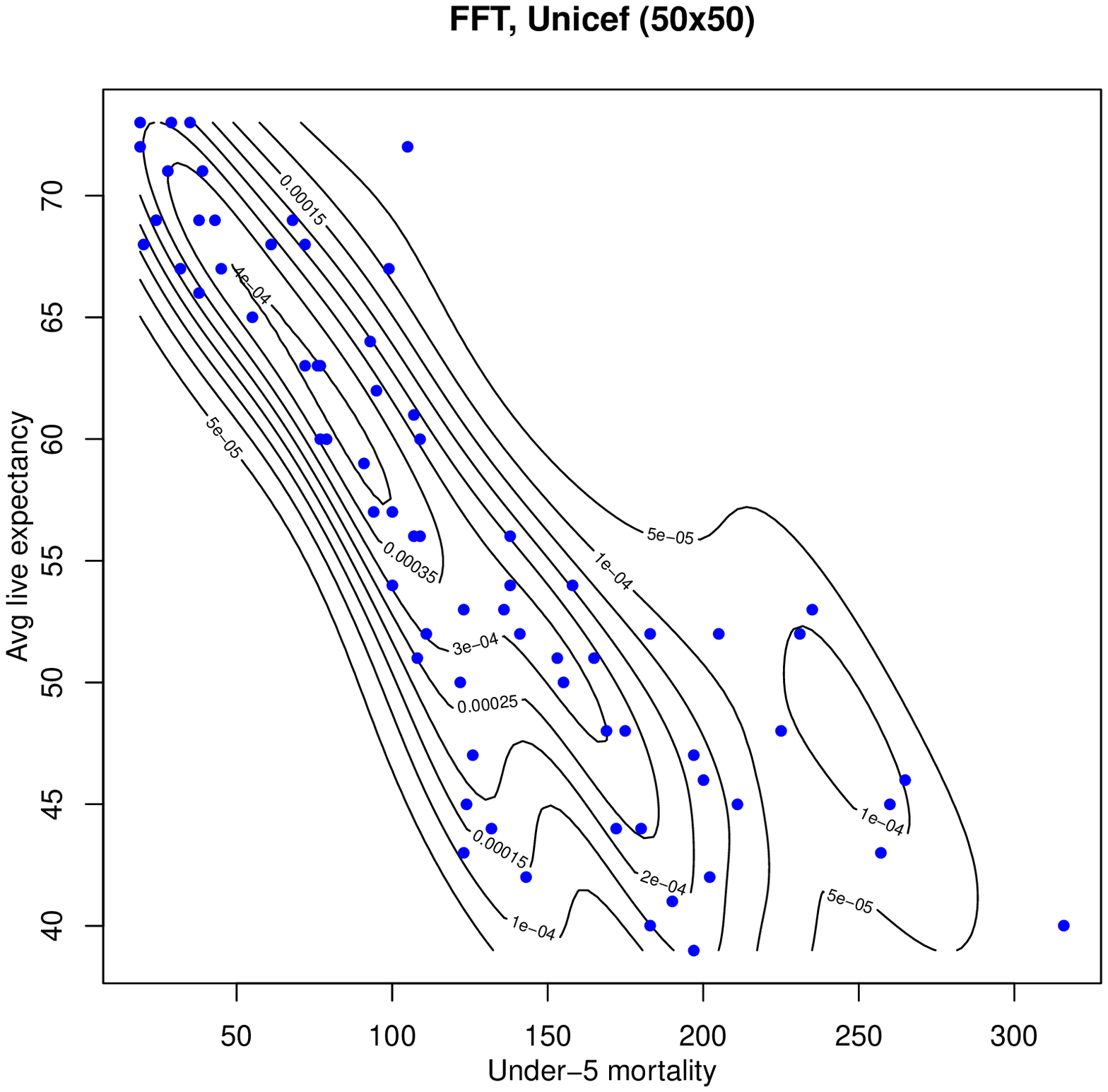} \\
   (e) \\  \vspace{0.0cm}
  \end{center}\end{minipage}
  \begin{minipage}[b]{0.5\textwidth}\begin{center}
   \includegraphics[scale=0.31]{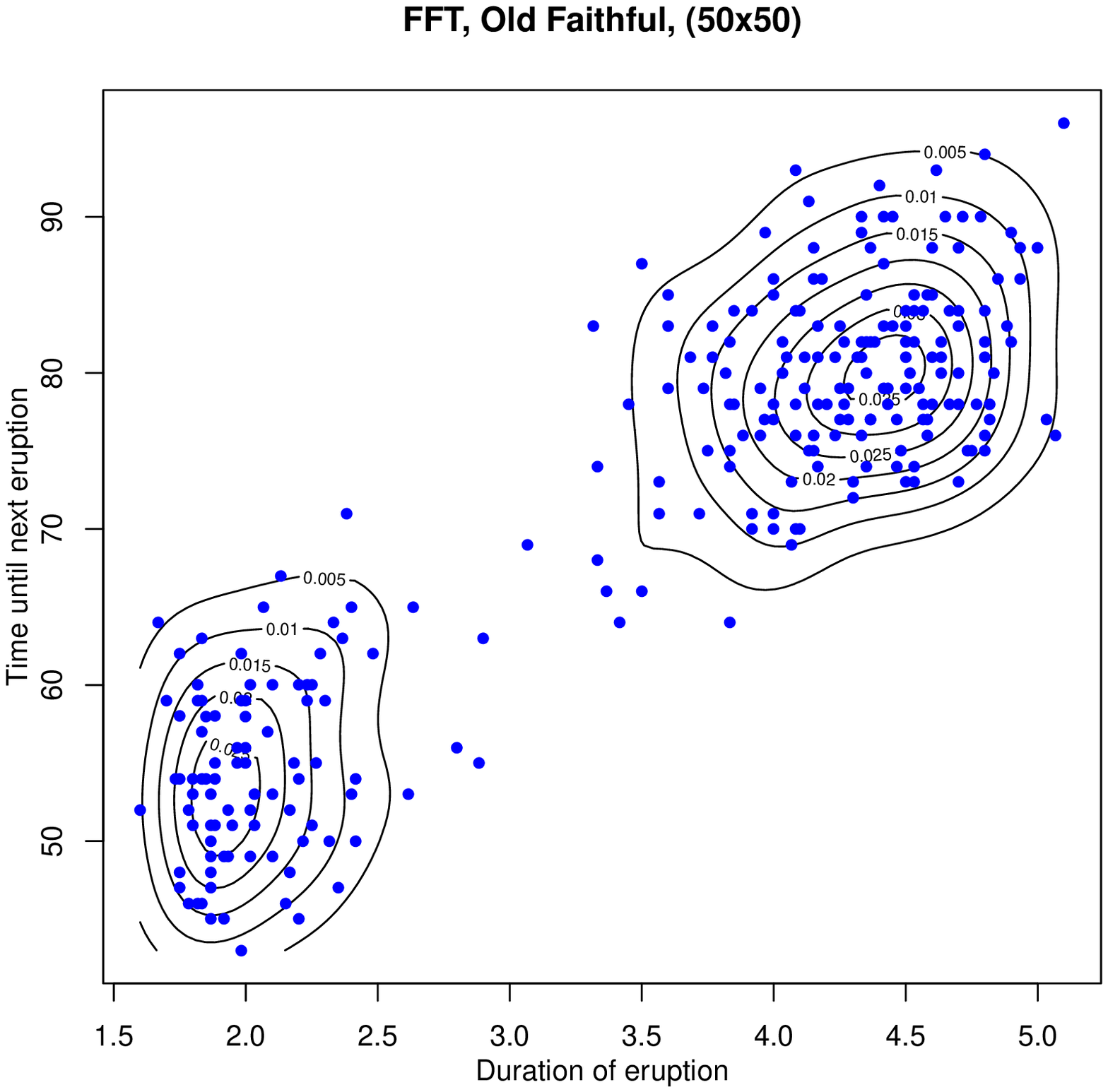} \\
   (f) \\  \vspace{0.0cm}
  \end{center}\end{minipage}    
  \caption{The effect of the binning and FFT procedures applied to \emph{Unicef} and \emph{Old Faithful} datasets. (a), (b) -- the optimal bandwidth based on Eq.~(\ref{eq-lscv-obj-fun-new}), that is if no binning is used, (c), (d) --  the optimal bandwidth based on Eq.~(\ref{eq-Psi-after-bin}), that is if binning is used and FFT is not used. Additionally, for better recognizing the effect of the binning, the plots from (a) and (b) (drawn with dotted lines) are added.  (e), (f) --  the optimal bandwidth based on Eq.~(\ref{eq-Psi-after-bin-2}) when FFT is used.} 
  	\label{fig-unicef-oldf}
\end{figure}

First we analyze how the binning procedure affects the accuracy of evaluating of the objective function $LSCV(\boldsymbol{H})$. In Figs. \ref{fig-unicef-oldf}(a) and \ref{fig-unicef-oldf}(b) we show densities of the Unicef and the Old Faithful datasets, respectively. The optimal bandwidth was calculated based on exact solution of the objective function given in Eq.~(\ref{eq-lscv-obj-fun-new}). In other words, no binning was used here. In Figs. \ref{fig-unicef-oldf}(c) and \ref{fig-unicef-oldf}(d) we can observe how the binning influences the resulting densities. Now the calculations were based on Eq.~(\ref{eq-Psi-after-bin}). As one can observe, even a moderate grid size (here $M_1=50, M_2=50$) is enough and the plots differ very slightly comparing with Figs. \ref{fig-unicef-oldf}(a) and \ref{fig-unicef-oldf}(b). After application of the FFT-based approach (this time calculations were based on Eq.~(\ref{eq-Psi-after-bin-2})) the resulting contour plots presented in Figs. \ref{fig-unicef-oldf}(e) and  \ref{fig-unicef-oldf}(f) are identical compared with those generated without the FFT-based support. This of course confirms the fact that the FFT-based procedure finds the correct bandwidth.

\subsection{Speed comparisons}\label{sec-speed-comparisions}
In this section we analyze how our FFT-based approach reduces the total computational times. Tree different implementations are considered. The first one is based on direct computation of the double summation in Eq.~(\ref{eq-Psi-before-bin}). This implementation is highly vectorized (no explicit \emph{for} loops; for details see supplementary material). We do not analyze a pure \emph{for}-loops-based implementation as it is extremely slow and, as such, is without any practical usability, especially for large $n$, like for example thousands or so. We called this implementation \emph{direct}. The second implementation utilizes the FFT and is based on Eq.~(\ref{eq-Psi-after-bin-2}), where precalculation of the kernel values (see Eq.~(\ref{eq-kj})) is vectorized.  We called this implementation \emph{fft-M}. Finally, the third implementation utilizes Eq.~(\ref{eq-Psi-after-bin-3}), that is a modified version of Eq.~(\ref{eq-Psi-after-bin-2}) where the sum limits $\{M_1, \ldots, M_d\}$ are replaced by some smaller values $\{L_1, \ldots, L_d\}$.  We called this implementation \emph{fft-L}.

In this experiment we do not find in fact the minimizer $\hat{\boldsymbol{H}}_{LSCV}$ of Eq.~(\ref{eq-H-minimizer}). This is because the different methods may require a different number of evaluations of the objective function under minimization. Instead of that, we measure execution times needed to calculate functionals defined in Eqs.~(\ref{eq-Psi-before-bin}), (\ref{eq-Psi-after-bin-2}) and (\ref{eq-Psi-after-bin-3}). In this experiment the time needed for binning is also included in the results (for \emph{fft-M} and \emph{fft-L} methods). Binning is a necessary step and as such should not be neglected.

To reduce the number of variants, all experiments were performed only for two-dimensional datasets. Additionally, in this experiment the statistical structure of the dataset is not very important, so the $\mathcal{N}(\boldsymbol{0}, \boldsymbol{I})$ distribution was used, varying only its size $n$. Using other distributions (i.e., these shown in Fig. \ref{fig-chacon-target-densities}) will not change the performance for the \emph{fft-M} implementation but can slightly change the performance for the \emph{fft-L} implementation. This is because some different $L_1$ and $L_2$ values may be assigned (see Eq.~(\ref{eq-L})) and, consequently, some different $P_1$ and $P_2$ values will be generated (see Eq.~(\ref{eq-Pk-2})). 

We took sample sizes from $n=200$ to $n=4000$ incrementing the sequence by $200$. Grid sizes are taken from $M_1=M_2=20$ to $M_1=M_2=200$ incrementing the sequence by $10$ (for simplicity, grids are equal in each direction, that is, $M_1=M_2$). For the \emph{fft-M} and \emph{fft-L} implementations each combination of the sample and grid sizes was used.  The computations were repeated 50 times and the mean time was calculated. For the \emph{direct} implementation 50 repetitions were computed for each sample size and also the mean time was calculated. 

\begin{figure}
	\centering
	\includegraphics[width=12.5cm]{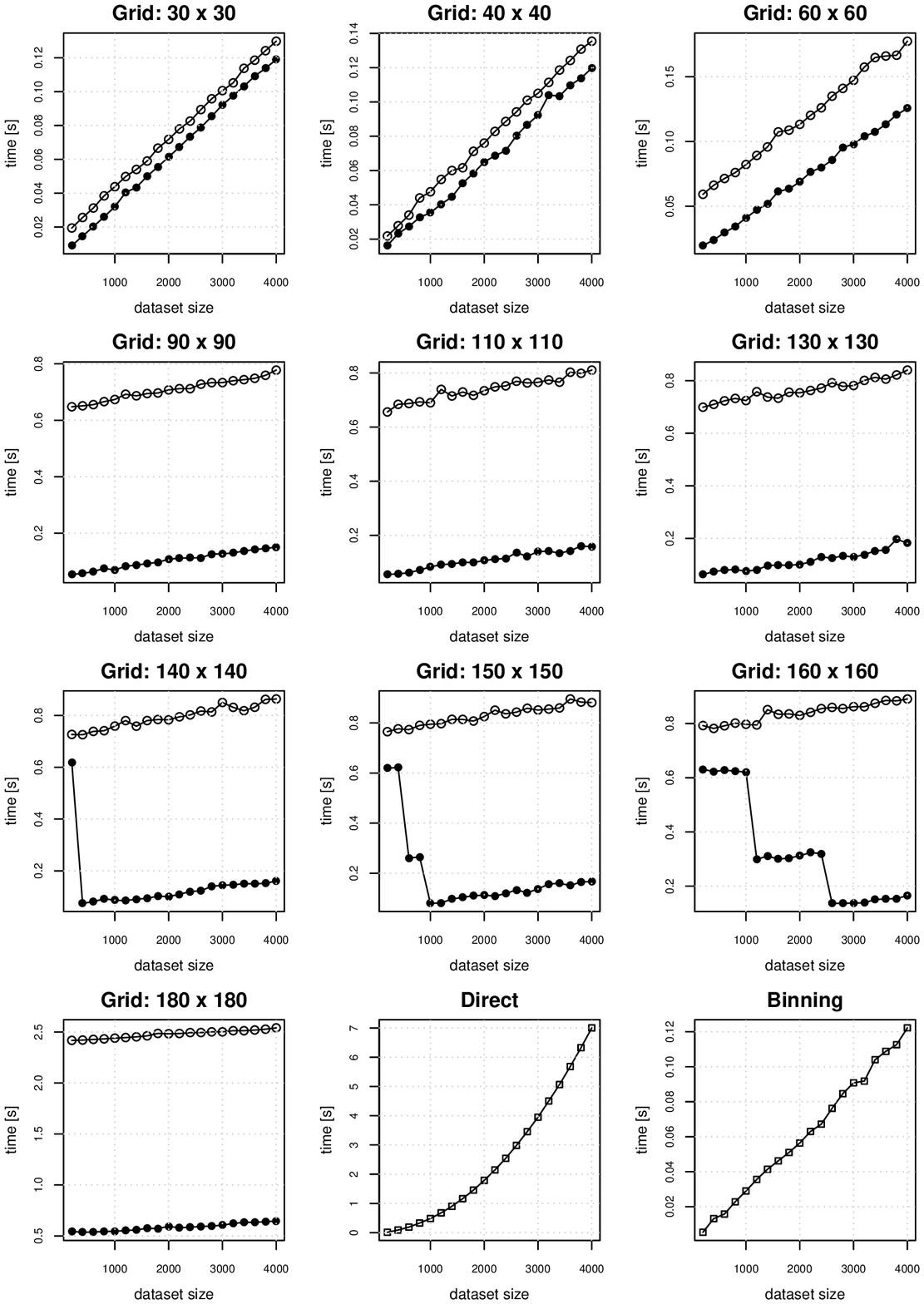}
	\caption{Speed comparison results for the \emph{fft-M}, \emph{fft-L} and \emph{direct} implementations. Lines marked with open circles ($\circ$) are for the \emph{fft-M} implementation, lines marked with filled circles ($\bullet$) are for the \emph{fft-L} implementation. The most lower-right plot gives result for the linear binning operation.}
	\label{fig-speed}
\end{figure}

In Fig. \ref{fig-speed} we show the results for the \emph{direct}, \emph{fft-M} and  \emph{fft-L} implementations. We can see that for small grid sizes (roughly up to $60 \times 60$) the calculation times are more dependent on the sample size compared with the grid sizes of about $90 \times 90$ and bigger. Starting from the grid sizes of  about $180 \times 180$, computational times become almost constant and this behavior is very attractive from the practical point of view. One could expect that the FFT-based implementations should not depend on $n$ (and this is true), but the main portion of the computational time for those small grid sizes comes from the binning step and not from the FFT calculations. As a consequence, we observe a linear dependence. If we remind ourselves that the binning is computed using a $O(n)$ algorithm, the observed linearity is obvious. 

Moreover, the \emph{fft-L} implementation is always faster compared with its \emph{fft-M} equivalent. The differences becomes bigger as the grid size increases. At the same time we can not see any significant accuracy degradation, so the usage of $L_k$ instead of $M_k$ is highly recommended in practical applications. 

We can also see an interesting behavior for the grid sizes $140 \times 140$, $150 \times 150$ and $160 \times 160$. Namely, computational times decrease as the sample size increases. The explanation of this phenomena is simple if we look carefully at Eq.~(\ref{eq-L}) and check for the values of $P_k$ which are calculated. Results for the three selected grid sizes are shown in Table \ref{tab-P}. For example, for the grid size $160 \times 160$ we can see that for sample sizes $n=\{200, 400, 600, 800, 1000\}$, $P_1$ and $P_2$ are both equal to $512$. Then, for sample sizes $n=\{1200, 1400, 1600, 1800, 2000, 2200, 2400\}$, $P_1$ and $P_2$ are equal to $256$ and $512$, respectively.  Finally, for the sample sizes $n=\{2600, 2800, 3000, 3200, 3400, 3600, 3800, 4000\}$,  $P_1$ and $P_2$ are both equal to $256$. The values of $P_k$ directly affect the FFT computation time (see Eq.~(\ref{eq-disc-conv-theo})), which cause the three `levels' in Fig. \ref{fig-speed} for the grid size $160 \times 160$. 

The last two plots in Fig.~\ref{fig-speed} confirm the $O(n^2)$ computational complexity of the \emph{direct} implementation and the $O(n)$ computational complexity of the binning operation. From a practical point of view, the usefulness of the \emph{direct} implementation is very controversial, especially for large data sizes.
\begin{table}
\centering
\caption{Values of $P_1$ and $P_2$ calculated according to Eq.~(\ref{eq-Pk-2}) for some selected grid and sample sizes.}
\label{tab-P}
\small
\begin{tabular}{ccccccccccc}
\hline
& \multicolumn{10}{c}{sample size $n$} \\
\cline{2-11}
\shortstack{grid\\size} & 200 & 400 & 600 & 800 & 1000 & 1200 & 1400 & 1600 & 1800 & 2000	\\
\hline
$140 \times 140$  & 
\shortstack{\\512\\512} & 
\shortstack{256\\256} &
\shortstack{256\\256} &
\shortstack{256\\256} &
\shortstack{256\\256} &
\shortstack{256\\256} &
\shortstack{256\\256} &
\shortstack{256\\256} &
\shortstack{256\\256} &
\shortstack{256\\256} \\
\hline
$150 \times 150$  & 
\shortstack{\\512\\512} & 
\shortstack{512\\512} & 
\shortstack{256\\512} &
\shortstack{256\\512} &
\shortstack{256\\256} &
\shortstack{256\\256} &
\shortstack{256\\256} &
\shortstack{256\\256} &
\shortstack{256\\256} &
\shortstack{256\\256} \\
\hline
$160 \times 160$  & 
\shortstack{\\512\\512} & 
\shortstack{512\\512} & 
\shortstack{512\\512} & 
\shortstack{512\\512} & 
\shortstack{512\\512} & 
\shortstack{256\\512} &
\shortstack{256\\512} &
\shortstack{256\\512} &
\shortstack{256\\512} &
\shortstack{256\\512} \\
\hline
& \multicolumn{10}{c}{sample size $n$} \\
\cline{2-11}
\shortstack{grid\\size} & 2200 & 2400 & 2600 & 2800 & 3000 & 3200 & 3400 & 3600 & 3800 & 4000 \\
\hline
$140 \times 140$  & 
\shortstack{\\256\\256} & 
\shortstack{256\\256} &
\shortstack{256\\256} &
\shortstack{256\\256} &
\shortstack{256\\256} &
\shortstack{256\\256} &
\shortstack{256\\256} &
\shortstack{256\\256} &
\shortstack{256\\256} &
\shortstack{256\\256} \\
\hline
$150 \times 150$  & 
\shortstack{\\256\\256} & 
\shortstack{256\\256} &
\shortstack{256\\256} &
\shortstack{256\\256} &
\shortstack{256\\256} &
\shortstack{256\\256} &
\shortstack{256\\256} &
\shortstack{256\\256} &
\shortstack{256\\256} &
\shortstack{256\\256} \\
\hline
$160 \times 160$  & 
\shortstack{\\256\\512} & 
\shortstack{256\\512} & 
\shortstack{256\\256} &
\shortstack{256\\256} &
\shortstack{256\\256} &
\shortstack{256\\256} &
\shortstack{256\\256} &
\shortstack{256\\256} &
\shortstack{256\\256} &
\shortstack{256\\256} \\
\hline
\end{tabular}
\end{table}

In Table \ref{tab-M-L} we show the values of $L_k$ calculated according to Eq.~(\ref{eq-L}) for some selected grid sizes $M_k$ and sample sizes $n$. As was expected, for a given value of the grid size, its equivalents $L_k$ are roughly constant, independently of the sample size. 
\begin{table}
\centering
\caption{Values of $L_k$ calculated according to Eq.~(\ref{eq-L}) for some selected  values of the grid and sample sizes, where $\tau=3.7$. Expressions in the parentheses mean $L_1$ and $L_2$ determined for given grid and sample sizes.}
\label{tab-M-L}
\small
\begin{tabular}{ccccccc}
\hline
& \multicolumn{6}{c}{grid size ($M = M_1 = M_2$)} \\
\cline{2-7}
\shortstack{sample\\size} & 30 & 60 & 110 & 140 & 160 & 200 \\
\hline
200 & $(14,13)$ & $(28,26)$ & $(52,47)$ & $(66,60)$ & $(75,68)$ & $(94,85)$ \\
600 & $(10,12)$ & $(19,24)$ & $(35,43)$ & $(44,55)$ & $(51,63)$ & $(63,78)$ \\
1000 & $(9,11)$ & $(19,21)$ & $(34,39)$ & $(43,50)$ & $(49,57)$ & $(61,71)$ \\
1400 & $(9,10)$ & $(18,20)$ & $(33,36)$ & $(41,46)$ & $(47,52)$ & $(59,65)$ \\
1800 & $(9,10)$ & $(18,19)$ & $(32,35)$ & $(41,45)$ & $(46,51)$ & $(58,64)$ \\
2200 & $(9,9)$ & $(17,18)$ & $(31,34)$ & $(40,43)$ & $(45,49)$ & $(57,61)$ \\
2600 & $(9,9)$ & $(17,18)$ & $(31,33)$ & $(39,42)$ & $(45,48)$ & $(56,60)$ \\
3000 & $(8,9)$ & $(17,18)$ & $(30,33)$ & $(39,41)$ & $(44,47)$ & $(55,59)$ \\
3400 & $(8,8)$ & $(17,16)$ & $(30,29)$ & $(38,37)$ & $(44,42)$ & $(55,53)$ \\
3800 & $(8,8)$ & $(16,16)$ & $(30,29)$ & $(38,37)$ & $(43,42)$ & $(54,52)$ \\
\hline
\end{tabular}
\end{table}

\subsection{V-statistics speed comparisons}\label{sec-eta-rs-speed-comparisions}
In this section we compare our FFT-based approach with some recent developments presented in \cite{Chacon:2015}. We have implemented the V-statistics given in Eq.~(\ref{eq-Qr}) using our FFT-based approach. Then we compare it with the \texttt{Qr.cumulant\{ks\}} R function. Note that the \texttt{Qr.cumulant\{ks\}} function (and other dependent functions, see supplementary material) has not been exported in the sense of the \textsf{R} environment's meaning. They are accessible only through the \emph{triple colon operator} \texttt{pkg:::name}. For safe, they should be treated as auxiliary ones. But anyway it seems that they work fine and return correct results.

In Eq.~(\ref{eq-Qr}) two issues should be considered separately. Firstly, how to efficiently calculate the $\eta_{r,s}$ entity and secondly, how to efficiently calculate the double sums there. The first issue has been successfully solved in the \texttt{ks} package and an efficient implementation is based on the \texttt{dmvnorm.deriv.recur\-sive\{ks\}} function. Our implementation of $\eta_{r,s}$ (and consequently FFT-based implementation of Eq.~(\ref{eq-V-statistics-degree-2})) is based on  this function. As for the second issue, in \cite{Chacon:2015} the authors also propose a method for fast calculation of the double sums but numerical simulations show that our FFT-based implementation is generally faster. To confirm this we have compared the \texttt{Qr.cumulant\{ks\}} achievements with our implementation (see the supplemental material for the up-to-date \textsf{R} source codes). We called the two implementations \emph{Qr-ks} and \emph{Qr-FFT}, respectively.

We took sample sizes from $n=100$ to $n=10000$ incrementing the sequence by $100$. Grid sizes are taken to be $M_1=M_2=50$ and $M_1=M_2=100$ (for simplicity, grids are equal in each direction, that is, $M_1=M_2$). Derivative orders are taken to be $r=\{0, 2, 4, 6, 8\}$. The computations were repeated 50 times and the mean time was calculated. To reduce the number of variants, all experiments were performed only for two-dimensional datasets. Additionally, in this experiment the statistical structure of the dataset is not very important, so the $\mathcal{N}(\boldsymbol{0}, \boldsymbol{I})$ distribution was used, varying only its size $n$. 

\begin{figure}[H]
  \begin{center}
   \includegraphics[width=11.3cm]{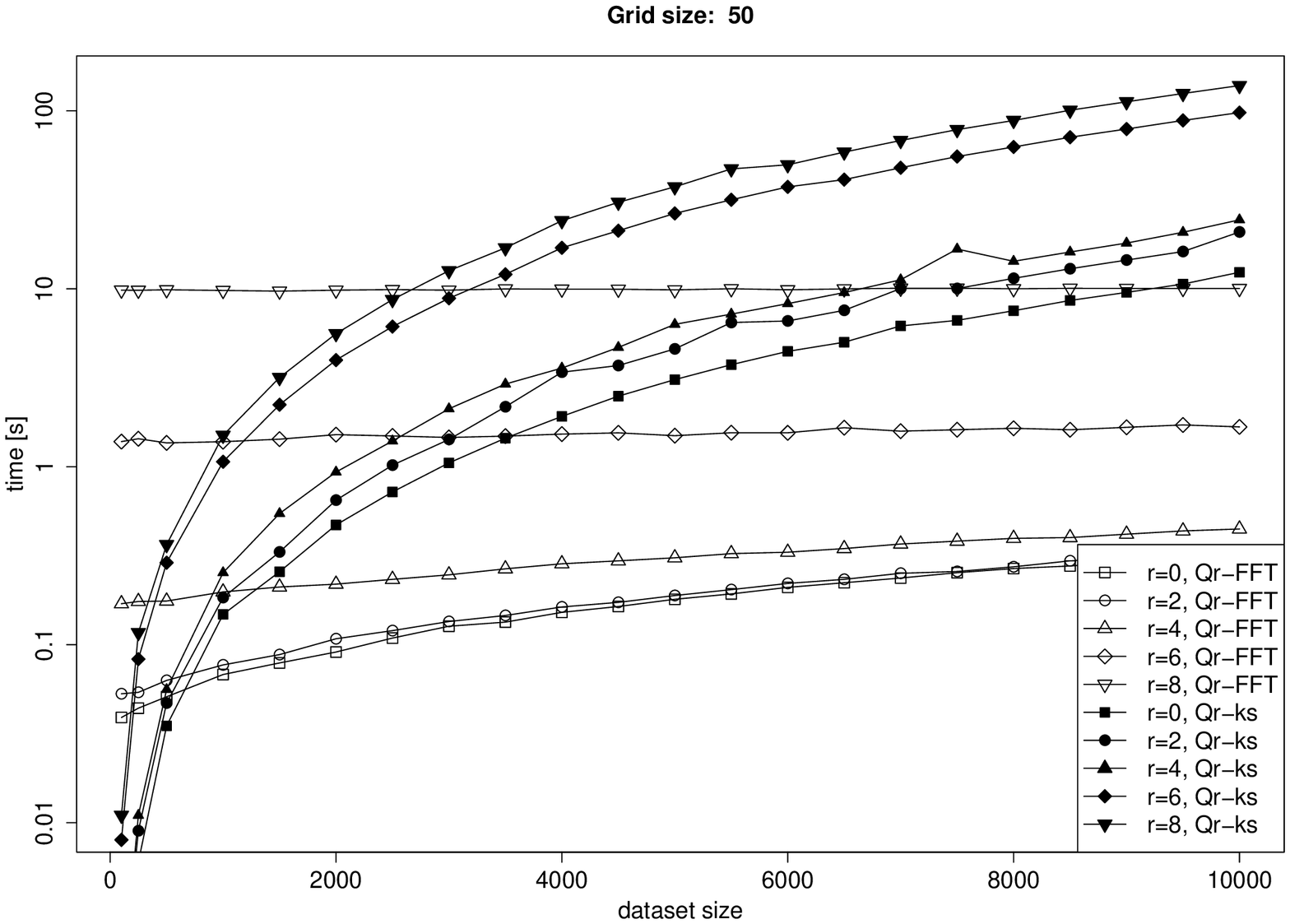} \\
   (a) \\ \vspace{2mm}
   \includegraphics[width=11.3cm]{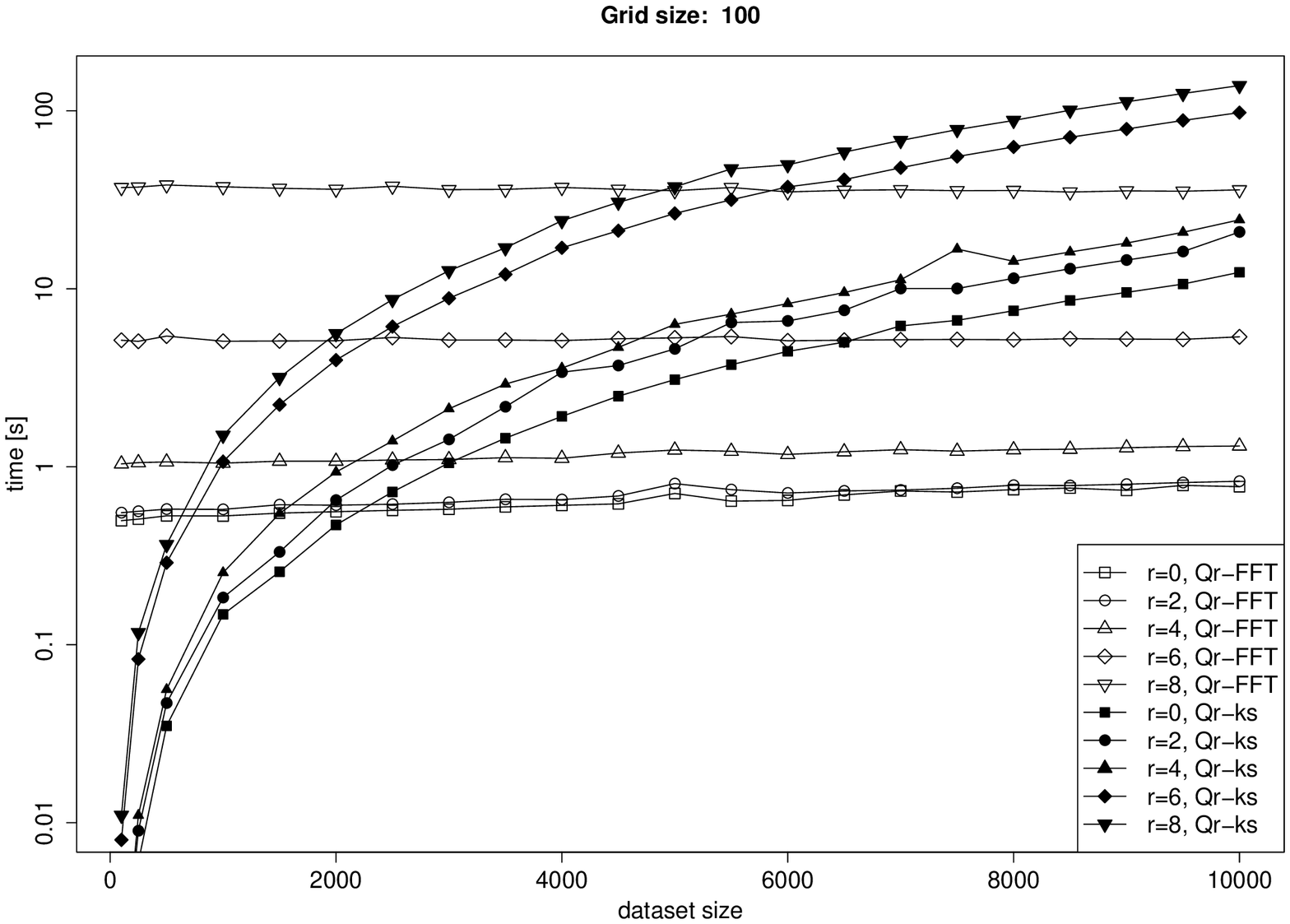} \\
   (b) \\ \vspace{2mm}
  \end{center}
  \caption{Speed comparison results for the \emph{Qr-ks} and the \emph{Qr-FFT} implementations. Lines marked with open symbols (like for example $\circ$) are for the \emph{Qr-FFT} implementation, lines marked with filled symbols (like for example $\bullet$) are for the \emph{Qr-ks} implementation.}
  \label{fig-speed-Qr}
\end{figure}

In Fig. \ref{fig-speed-Qr} we show the results for both the \emph{Qr-ks} and the \emph{Qr-FFT} implementations. As can be seen, the \emph{Qr-FFT} implementation usually outperforms the \emph{Qr-ks} implementation, especially for bigger $n$. The \emph{Qr-ks} implementation is faster only for smaller $n$. What is also important, the \emph{Qr-FFT} implementation is practically independent of the sample size. Some small departures from this behavior (visible mainly for grid size equals $50$ and for smaller derivative orders) are due to the binning (which is the required data preprocessing step). As the binning takes relatively not so much time, its influence for the total computational time for bigger $n$ and $r$ can be neglected. 

\section{Conclusion}\label{sec-conclusion}
Although nonparametric kernel density estimation is nowadays a standard technique in exploratory data analyses, there still exist some not fully solved problems. They include among other things, the question of how to fast compute the bandwidth, especially for the multivariate unconstrained case. In the paper we have presented a satisfactory FFT-based algorithm which is fast, accurate and covers unconstrained bandwidth matrices. We have outlined a complete procedure and have made comprehensive simulation studies. 

Our main contributions in the paper are: (a) paying attention to a real problem involving the FFT in the field of bandwidth selection, and (b) improving the existing FFT-base algorithm by proposing a new reshaping shown in Eqs. (\ref{eq-K-2D-new}) and (\ref{eq-C-2D-new}). The latter plays a  crucial role in adapting the FFT-based algorithm for supporting both constrained and unconstrained bandwidth matrices. (c) pointing out how our algorithm can be used for fast computation of integrated density derivative functionals involving an arbitrary derivative order. This is extremely important in implementing almost all modern bandwidth selection algorithms, not only the LSCV one used in the paper. Preliminary computer simulations proves the practical usability of our approach.

In Section \ref{sec-synth-data} we have reported some numerical-like problems which remain a challenging open problem. 

\section*{Acknowledgments}
The authors would like to thank anonymous reviewers whose comments greatly helped improve the quality of the
manuscript. 

\section*{References}

\bibliographystyle{elsarticle-num} 
\bibliography{CSDA_GRAMACKI}

\end{document}